\documentclass[aps, pra, reprint, superscriptaddress]{revtex4-1}
\usepackage{latexsym}
\usepackage{amsthm}
\usepackage{amssymb}
\usepackage{amsmath}
\usepackage[all]{xy}
\usepackage{float}
\usepackage{graphicx}
\usepackage{subfigure}
\usepackage{dcolumn}
\usepackage{bm}
\usepackage{bbm}
\usepackage[usenames, dvipsnames]{xcolor}
\usepackage{amsfonts}
\usepackage{cases}
\usepackage[a4paper, pagebackref=false, colorlinks=true,
linkcolor=blue, citecolor=blue,
pdfauthor={ },
pdftitle={ },
pdfsubject={ },
pdfkeywords={ }]{hyperref}

\begin{document}

\title{Search for ultralight dark matter with a frequency adjustable diamagnetic levitated sensor}

\author{Rui Li}
\affiliation{CAS Key Laboratory of Microscale Magnetic Resonance and School of Physical Sciences, University of Science and Technology of China, Hefei 230026, China}
\affiliation{CAS Center for Excellence in Quantum Information and Quantum Physics, University of Science and Technology of China, Hefei 230026, China}

\author{Shaochun Lin}
\author{Liang Zhang}
\affiliation{CAS Key Laboratory of Microscale Magnetic Resonance and School of Physical Sciences, University of Science and Technology of China, Hefei 230026, China}
\affiliation{CAS Center for Excellence in Quantum Information and Quantum Physics, University of Science and Technology of China, Hefei 230026, China}

\author{Changkui Duan}
\affiliation{CAS Key Laboratory of Microscale Magnetic Resonance and School of Physical Sciences, University of Science and Technology of China, Hefei 230026, China}
\affiliation{CAS Center for Excellence in Quantum Information and Quantum Physics, University of Science and Technology of China, Hefei 230026, China}

\author{Pu Huang}
\email{hp@nju.edu.cn}
\affiliation{National Laboratory of Solid State Microstructures and Department of Physics, Nanjing University, Nanjing, 210093, China}

\author{Jiangfeng Du}
\affiliation{CAS Key Laboratory of Microscale Magnetic Resonance and School of Physical Sciences, University of Science and Technology of China, Hefei 230026, China}
\affiliation{CAS Center for Excellence in Quantum Information and Quantum Physics, University of Science and Technology of China, Hefei 230026, China}

\begin{abstract}	
Among several dark matter candidates, bosonic ultra-light (sub-meV) dark matter is well motivated because it could couple to the Standard Model (SM) and induce new forces. Previous MICROSCOPE and Eöt-Wash torsion experiments have achieved high accuracy in the sub-1 Hz region, but at higher frequencies there is still a lack of relevant experimental research. We propose an experimental scheme based on the diamagnetic levitated micromechanical oscillator, one of the most sensitive sensors for acceleration sensitivity below the kilohertz scale. In order to improve the measurement range, we used the sensor whose resonance frequency $\omega_0$ could be adjusted from 0.1Hz to 100Hz. The limits of the coupling constant $g_{\scriptscriptstyle B-L}$ are improved by more than 10 times compared to previous reports, and it may be possible to achieve higher accuracy by using the array of sensors in the future.

\end{abstract}
\maketitle

\section{ INTRODUCTION}
There are many astronomical \cite{doi:10.1146/annurev.astro.39.1.137, Massey_2010} and cosmological observations  \cite{Markevitch_2004} that prove the existence of dark matter particles\cite{bertone_2010,Salucci_2019}, but the specific parameters of dark matter, especially the quality, are still highly uncertain \cite{PhysRevD.98.030001}. Many direct detection studies have assumed that dark matter is composed of supersymmetric fermions, but so far there has not been enough evidence. Now the focus of research is gradually shifting to ultralight bosons and the quality range is approximately $10^{-22} \mathrm{eV}\textless m_{\phi}\textless0.1\mathrm{eV}$ \cite{IRASTORZA201889, PhysRevD.95.043541}. For ultralight bosons with a mass less than 1eV, due to their high particle number density, they behave like a classical field. Due to the viral theorem , if the DM has virialized to the Galaxy, it will be moving with a typical speed $v_{\scriptscriptstyle \mathrm{DM}} \approx 10^5 $m/s \cite{Bovy_2012, PhysRevD.98.103006,10.1093/mnras/stx3262}. This corresponds to Compton frequency $\omega_{s}=m_{\phi}/ \hbar $ and De Broglie wavelength $\lambda_{\scriptscriptstyle \mathrm{DM}}=hc^2/(m_{\phi} v_{\scriptscriptstyle \mathrm{DM}})$.

According to the previous reports, such as ADMX \cite{PhysRevLett.120.151301} can search for the Peccei-Quinn axion in the mass range $10^{-6}\mathrm{eV}\textless m_{\phi}\textless 10^{-3} \mathrm{eV}$ \cite{backes2021quantum,PhysRevLett.126.191802}. And the pseudoscalar axion-like ULMBs with masses between $10^{-23}\mathrm{eV} $ and $10^{-18}\mathrm{eV}$ \cite{PhysRevX.7.041034,PhysRevLett.122.231301,smorra2019direct} and scalar dilaton ULMBs with masses between $10^{-21}$eV and $10^{-5}\mathrm{eV}$ by use ultrastable clocks \cite{PhysRevLett.125.201302,PhysRevD.91.015015} and gravitation wave detectors \cite{vermeulen2021direct}
have recently been reported.

When DM is a vector field couples to a conserved current, corresponding to the baryon number minus lepton number (B$-$L charge) in the SM. The Lagrangian in this case can be written as \cite{Carney_2021}:
\begin{equation}\label{Lagra}
\mathcal{L}=-\frac{1}{4} F_{\mu \nu} F^{\mu \nu} -\frac{1}{2} m_{\phi}^2 A^2 +i g_{\scriptscriptstyle B-L} A_{\mu} \overline{n} \gamma^{\mu} n
\end{equation}
where $n$ is the neutron field and the DM field couples directly to the number of neutrons, $g_{\scriptscriptstyle B-L}$  is the coupling strength.
Using the Lorentz gauge and the plane wave approximation, the dark electric field can be written as: $E\approx\sqrt{\rho_{\scriptscriptstyle \mathrm{DM}}} \mathrm{sin} (\omega_s t-\vec {k} \cdot \vec{x})$, where $\rho_{\scriptscriptstyle \mathrm{DM}}\approx 0.3\mathrm{GeV}/\mathrm{cm^3}$  \cite{Read_2014} is the local DM density.

In ground experiments, assume that using a magnet-gravity mechanical oscillator to measure the ultralight DM field along the Earth's axis,  we can parameterize the force exerted on the sensor as:
\begin{equation}\label{Fsig}
F_{\mathrm{sig}}(t)=\alpha g_{\scriptscriptstyle B-L}  N_g F_0  \mathrm{sin}(\omega_s t)
\end{equation}

\noindent because the De Broglie wavelength of DM is much larger than the size of the sensor so that we drop the $x$ dependence. In this equation, $\alpha=\mathrm{sin} \theta_N$ denotes the component along the direction of gravity and $\theta_N$ means the latitude of the location of the ground experiment system.  In order to avoid the effects of the Earth's rotation under long time measurements and increase the force, experiment system is best carried out at high latitudes like in the Arctic which $\alpha=1$. $F_0=\sqrt{\rho_{\scriptscriptstyle \mathrm{DM}}}\approx 10^{-15}$N and $N_g$ is the total number of neutrons  in the sensor. We can approximate write it as $N_g\approx \frac{1}{2} m/m_{\mathrm{neu}}$ in a sensor with mass $m$ and $m_{\mathrm{neu}}$ is the neutron mass.  The force $F_{\mathrm{sig}}(t)$ is proportional to the mass of the sensor, 
so the main criterion about the sensor is acceleration sensitivity.

Here we propose a experiment scheme to detect DM using a frequency adjustable diamagnetic levitated sensor. The resonance frequency could be changed by adjust the magnetic field gradient in a paramagnetic part of the oscillator and frequency range from 0.1Hz to 100Hz.
 This means that we have high detection accuracy to detect DM with mass in the range from $10^{-16}$eV to $10^{-13}$eV. 
 Compare to previously reported experiments, our experiment scheme can achieve more than one order of magnitude improvement in the measurement of the coupling strength $g_{\scriptscriptstyle B-L}$ based on the results of theoretical calculation.

\section{THEORETICAL CALCULATION}
 
 Under the effect of the ultralight DM field, consider thermal noise and measurement noise,
 the motion equation of a mechanical oscillator at resonant frequency $\omega_0$ could be written as: 
\begin{equation}
	m\ddot{x}+ m\gamma \dot{x} + m\omega_0^2 x 
	=F_{\mathrm{sig}}(t)+F_{\mathrm{th}}+F_{\mathrm{mea}}
\end{equation}
where $\gamma$ is damp coefficient; 
the $F_{\mathrm{sig}}(t)$ is the DM field drive from equation (\ref{Fsig}); $F_{\mathrm{th}}$ is the environmental thermal noise; and the $F_{\mathrm{mea}}$ represents the measurement noise which is mainly composed of  the detector imprecision noise and backaction of radiation pressure fluctuations.

The total acceleration noise of the system is given by:
\begin{equation}\label{Saatot}
	S_{\mathrm{aa}}^{\mathrm{tot}}= S_{\mathrm{aa}}^{\mathrm{th}}+	(\frac{S_{\mathrm{xx}}^{\mathrm{imp}}}{|\chi_{\mathrm{\scriptscriptstyle m}}(\omega,\omega_0)|^2}+ \frac{S_{\mathrm{ff}}^{\mathrm{ba}}}{m^2} )
\end{equation}
where $\chi_{\mathrm{\scriptscriptstyle m}}(\omega,\omega_0)$ is the mechanical susceptibility given by $|\chi_{\mathrm{\scriptscriptstyle m}}(\omega,\omega_0)|^2=1/[(\omega^2-\omega_0^2)^2+\gamma^2 \omega^2]$, %$S_{aa}^{sig} $ is the acceleration PSD of DM field which we want to detect,
and $S_{\mathrm{aa}}^{\mathrm{th}} =4 \gamma k_B \mathrm{T}/m $ is the thermal noise where $k_B$ is Boltzmann constant and T indicates environment temperature.
The detector imprecision noise $S_{\mathrm{xx}}^{\mathrm{imp}}$ and the backaction noise $S_{\mathrm{ff}}^{\mathrm{ba}}$ 
make up the total measurement noise 
$S_{\mathrm{aa}}^{\mathrm{mea}}=S_{\mathrm{xx}}^{\mathrm{imp}} /|\chi_{\mathrm{\scriptscriptstyle m}}(\omega,\omega_0)|^2 +S_{\mathrm{ff}}^{\mathrm{ba}} / m^2$, 
and $S_{\mathrm{xx}}^{\mathrm{imp}}\cdot S_{\mathrm{ff}}^{\mathrm{ba}}=(1/\eta) \hbar^2$ meanwhile. 
Here $\eta\leqslant 1$ is the measurement efficiency, and $\eta= 1$ corresponding to standard quantum limit (SQL).

The total measurement noise $S_{\mathrm{aa}}^{\mathrm{mea}}$ for the sensor operating at SQL condition at resonance frequency $\omega_0$ could be given by the simple  formula \cite{Clark2016SidebandCB}:
 \begin{equation}
 	S_{\mathrm{aa}}^{\mathrm{mea,SQL}}=\frac{2 \sqrt{(\omega_0^2-\omega^2)^2+\gamma^2 
 			\omega^2}}{m}
 \end{equation}
 And achieving the SQL in a frequency range need to optimize the measurement parameters 
  frequency by frequency as the range is scanned.

\begin{figure}[t]
   	\centering
   	\includegraphics[width=0.95\columnwidth]{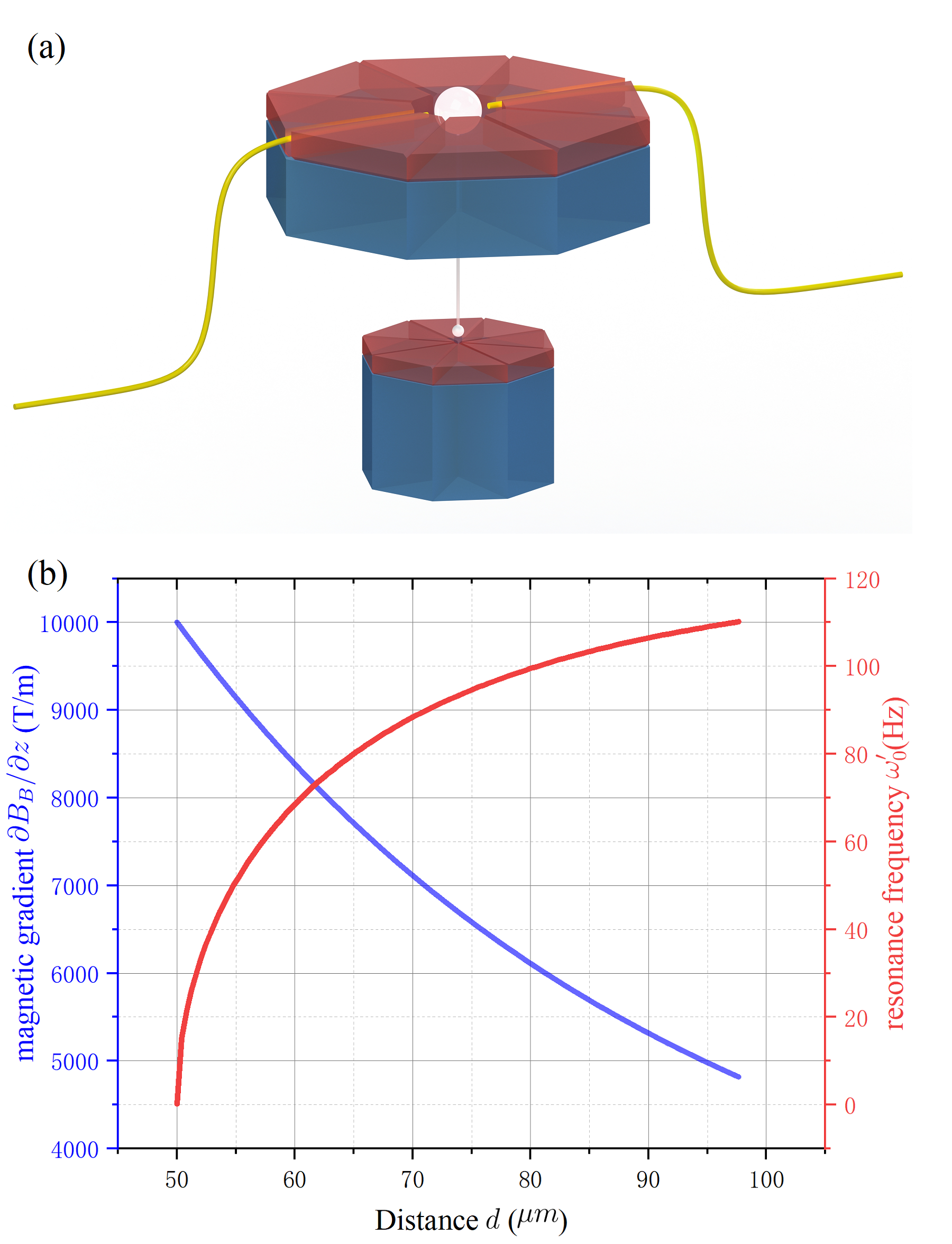}
   	\caption{(a) Schematic diagram of the experimental setup. A diamagnetic sphere of 0.5 mm radius is levitated in the magnetic gravity trap, and a paramagnetic  microsphere of 11 $\mu m$ radius	is connected to the upper diamagnetic sphere by a thin glass rod. A 1550 nm laser is transmitted through the left fibre to the right fibre, passing the transparent diamagnetic sphere. 
   		(b) The magnetic field gradient $\partial B_{\scriptscriptstyle B}/\partial z$ and the resonance frequency $\omega_0^{\prime}$ changes by the relative distance $d$, expressed by the blue and red lines respectively.}
   	\label{fig1}
   \end{figure}
   
We use the total acceleratioon noise $S_{\mathrm{aa}}^{\mathrm{tot}}$ as the acceleration measurement sensitivity of the system. From the equations (\ref{Fsig})-(\ref{Saatot}), consider the optimal case of $\alpha=1$, we obtain the relationship between coupling strength $g_{\scriptscriptstyle B-L}$  and the acceleration measurement sensitivity $S_{\mathrm{aa}}^{\mathrm{tot}}$ by: 
 \begin{equation}
 	\label{gBL}
 	g_{\scriptscriptstyle B-L}= \frac{2 m_{neu}}{F_0}  \sqrt{\frac{S_{\mathrm{aa}}^{\mathrm{tot}}}{T_{\mathrm{tot}}}}
 \end{equation}
 where $T_{\mathrm{tot}}$ denotes the effective total integration time. The DM signal is essentianlly a coherent force and the timescales $T_{\mathrm{coh}} \approx 10^6/ \omega_s$.
 When the DM frequency $\omega_s$ is lower to satisfy $ T_{\mathrm{coh}}\textgreater T_{\mathrm{mea}}$, 
  all the measurement time $T_{\mathrm{mea}}$ contributes to the coherent DM signal. And as the DM frequency $\omega_s$  increases, when $ T_{\mathrm{coh}}\textless T_{\mathrm{mea}}$, only the proportion of $T_{\mathrm{coh}}/T_{\mathrm{mea}}$ in the measurement time contributes to the coherent signal. So we  define the effective integration time:
 \begin{equation}
 	T_{\mathrm{tot}}=\left\{ \begin{array}{ll}
 		T_{\mathrm{mea}} & \textrm{if $T_{\mathrm{coh}}< T_{\mathrm{mea}}$}\\
 		\sqrt{T_{\mathrm{mea}} \cdot T_{\mathrm{coh}}} & \textrm{if $T_{\mathrm{coh}}>  
 			T_{\mathrm{mea}}$} \notag
 	\end{array} \right. 
 \end{equation}

 \section{EXPERIMENTAL SCHEME}
 The levitated micromechanical and nanomechanical oscillators have been demonstrated as one of the ultrasensitive acceleration sensors due to its ultralow dissipation \cite{PhysRevApplied.15.024061,PhysRevA.101.053835}. 
We propose a reasonable scheme by our calculation as shown in  Fig.\ref{fig1}(a). A diamagnetic sphere made by PMMA with radius $r_1$=0.5mm(corresponding volume $V_1$), density $\rho_1$ and magnetic susceptibility $\chi_{\scriptscriptstyle 1}$  is levitated in the upper magnet (name as $Magnet\text{-}A$) center region, and the oscillator signal is detected through the fibre on both sides. 
 A paramagnetic microsphere made by $\mathrm{Tb_2 O_3}$ with 
 radius $r_2=11 \mu$m(corresponding volume $V_2$), density $\rho_2$ and magnetic susceptibility $\chi_{\scriptscriptstyle 2}$ is connected to the upper diamagnetic sphere through a thin glass rod.  And another combined magnets (name as $Magnet\text{-}B$) is placed under the paramagnetic microsphere. The whole magnet assembly is placed in a multi-stage suspension system, and uses active vibration isolation devices to further improve the isolation 
 effect\cite{ACERNESE2010182,Acernese_2015}.

   \begin{figure*}[t]
  	\centering
  	\includegraphics[width=1.92\columnwidth]{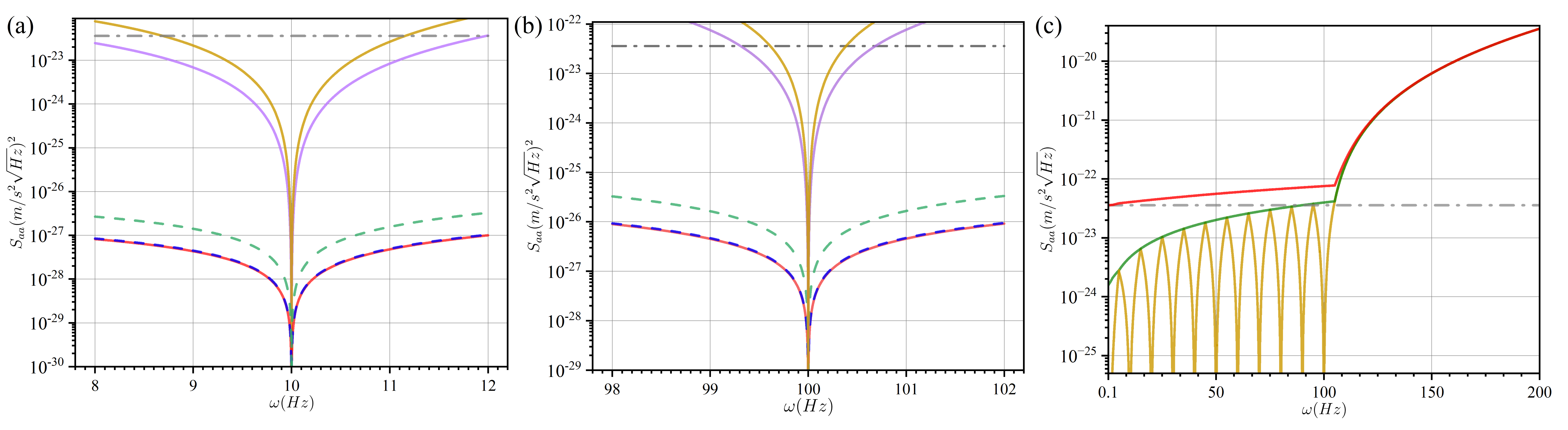}
  	\caption{Acceleration power spectral density $S_{\mathrm{aa}}$ 
  		(a) Resonance frequency $\omega_0$=10Hz, the grey dashed line indicates the thermal noise $S_{\mathrm{aa}}^{\mathrm{th}}$; 
  		the red line indicates the acceleration detection noise $S_{\mathrm{aa}}^{\mathrm{mea,SQL}}$;
  		the blue dashed line indicates the $S_{\mathrm{aa}}^{\mathrm{mea}}$ with the optimal light intensity $P_{\mathrm{opt}}(\omega,\omega_0)$ in each frequency between 8Hz to 12Hz and the measurement efficiency $\eta$=1; 
  		the green dashed line indicates the same $P_{\mathrm{opt}}(\omega,\omega_0)$ as the blue dashed line but the $\eta$=0.1; 
  		the purple line indicates the light intensity $P_{\mathrm{opt}}(\omega_0,\omega_0)$ and $\eta$=1; 
  		the yellow line indicates the same $P_{\mathrm{opt}}(\omega_0,\omega_0)$ and $\eta$=0.1; 
  		(b) Resonance frequency $\omega_0$=100Hz, and the others are the same as (a); 
  		(c) Adjust resonance frequency $\omega_0$ from 0.1Hz to 100Hz, the grey dashed line indicates the thermal noise $S_{\mathrm{aa}}^{\mathrm{th}}$;
  		the yellow line indicates the acceleration measurement noise $S_{\mathrm{aa}}^{\mathrm{mea}}$ with $\eta$=0.1, and here the scan step $\Delta \omega_s$=10Hz it is only used to show the measurement scheme;
  		the green line indicates the envelope of the yellow line in the diagram and write it as $S_{\mathrm{aa}}^{\mathrm{mea}^{\prime}}$; 
  		the red line is the acceleration measurement sensitivity $S_{\mathrm{aa}}^{\mathrm{tot}}=S_{\mathrm{aa}}^{\mathrm{th}}+S_{\mathrm{aa}}^{\mathrm{mea}^{\prime}}$.}\label{fig3}
  \end{figure*}
  
$Magnet\text{-}A$ is constructed in a similar way to our previous articles\cite{PhysRevResearch.2.013057}. And need to use high remanence magnetic material with two different magnetisation direction to generate enough magnetic force. The red  express the direction point to the centre, and the blue express the direction out to the centre. In addition, using a less remanence magnetic material to build the upper layer of $Magnet\text{-}B$ and high magnetic material to build the lower layer. The combination of two different remanence magnetic materials allows $Magnet\text{-}B$ to have a higher magnetic field gradient while reducing the magnetic field strength. And the direction of magnetisation is also indicated by red and blue colours.

The magnetic field energy of the upper paramagnetic sphere can be written as:
 \begin{equation}
 U_1=-\int_{V_1}\frac{\chi_{\scriptscriptstyle 1}}{2\mu_0}  B_{\scriptscriptstyle \mathrm{A}} ^2 dV
\end{equation}
 where $B_{\scriptscriptstyle \mathrm{A}} $ represents the magnetic field created by
  $Magnet\text{-}A$.
 Assuming that the $Magnet\text{-}B$ is far away at beginning , the $z$ direction equilibrium position $z_0$ of the oscillator in the magnetic-gravity trap satisfies:
$\partial U_1/\partial z |_{z=z_0}=(\rho_1 V_1+\rho_2 V_2 )g$.
 And the resonance frequency in $z$ direction is:
  \begin{equation}\label{omega_0}
\omega_0=\sqrt{\frac{1}{\rho_1 V_1+\rho_2 V_2}\cdot
 	\frac{\partial^2 U_1}{\partial z^2}}\bigg|_{z=z_0}
\end{equation}

Then we make the $Magnet\text{-}B$ rise, the magnetic field $B_{\scriptscriptstyle B} $ from $Magnet\text{-}B$ in the lower paramagnetic microsphere will become  larger. And because of $V_2\ll V_1$, we  can simplify the magnetic field energy of the paramagnetic microspheres as  $U_2=-\chi_{\scriptscriptstyle 2} B_{\scriptscriptstyle \mathrm{B}}^2 V_2/2\mu_0$.
Now the resonance frequency along $z$ direction of the oscillator change as:
 \begin{equation}\label{omega_B/z}
\omega_0^{\prime}=\sqrt{\omega_0^2-\frac{\chi_{\scriptscriptstyle 2}V_2}{\mu_0(\rho_1 
V_1+\rho_2V_2)} \left( \frac{\partial B_{\scriptscriptstyle B}}{\partial z} \right)^2}\bigg|_{z=z_0}
  \end{equation}
 where $ \chi_{ \scriptscriptstyle 2}\textgreater 0$ and $\omega_0^{\prime} \textless \omega_0$.
We ignore the second order gradient term because of
  $(\partial B_{\scriptscriptstyle \mathrm{B}}/\partial z)^2\gg B_{\scriptscriptstyle \mathrm{B}}  (\partial^2 B_{\scriptscriptstyle B} / \partial z^2) $.
And the magnetic force from $Magnet\text{-} B$ on the paramagnetic microsphere is much lower than the total gravity of oscillator since $B_{\scriptscriptstyle \mathrm{B}}$ and $V_2$  are very small, the equilibrium position $z_0$ will not be changed therefore.
 
 We use finite element method to simulate the magnetic field gradient $\partial B_{\scriptscriptstyle \mathrm{B}}/\partial z$ changes by the distance between the  paramagnetic microsphere and $Magnet\text{-}B$ expressed by $d$ range from 50$\mu$m to 100 $\mu$m, then use equation (\ref{omega_B/z}) to calculate the corresponding resonance  frequency $\omega_0^{\prime}$, as shown in  Fig.\ref{fig1}(b). It is theoretically possible to bring the resonance frequency  $\omega_0^{\prime}$ close to zero by reducing the distance $d$. But in order to improve the stability of the oscillator and reduce the requirement for the isolation system, we select resonance frequency $\omega_0^{\prime}$ variation range from 0.1Hz to 100Hz.

 \section{EXPERIMENTAL RESULT ESTIMATE}
 
 Now we calculate the acceleration measurement sensitivity of this system. In order to improve the acceleration sensitivity, the whole system was placed  in a low temperature environment which T=30mK, and estimate the damp coefficient $\gamma=10^{-4}$Hz \cite{PhysRevApplied.15.024061,PhysRevApplied.16.L011003}. In the Supplementary material, we calculate the dependence of the total measurement noise $S_{\mathrm{aa}}^{\mathrm{mea}}$ on the laser input power $P_{\mathrm{in}}$ and obtained the optimized laser input 
 power $P_{\mathrm{opt}}(\omega,\omega_0)$ to minimised the total measurement noise.

 In the cases of the oscillator resonance frequency $\omega_0$ equal to 10Hz and 100Hz,
we calculate the corresponding acceleration noise  and the results are shown in Fig.\ref{fig3}(a) and Fig.\ref{fig3}(b). When resonance frequency $\omega_0=10$Hz, 
assuming measurement efficiency $\eta=1$ and we set the laser input power to optimal laser power for each point as $P_{\mathrm{opt}}(\omega,\omega_0)$, the measurement noise $S_{\mathrm{aa}}^{\mathrm{mea}}$  can almost reach the SQL at this time. 
With the measurement efficiency $\eta$ reduce to 0.1, the measurement noise is slightly increased. 
But actually, to simplify the experiment, the laser input power need to choose near the resonance frequency $\omega_0$ by $P_{\mathrm{opt}}(\omega_0,\omega_0)$, it will make the measurement noise $S_{\mathrm{aa}}^{\mathrm{mea}}$ increase rapidly. 
In Fig.\ref{fig3}(a), in the frequency range from 9Hz to 11Hz, the measurement noise $S_{\mathrm{aa}}^{\mathrm{mea}}$ is always below the thermal noise $S_{\mathrm{aa}}^{\mathrm{th}}$ with $\eta=0.1$. When the resonance frequency $\omega_0$ is adjusted to 100Hz, the range of measurement noise $S_{\mathrm{aa}}^{\mathrm{mea}}$ below thermal noise $S_{\mathrm{aa}}^{\mathrm{th}}$ is reduced to 99.6Hz to 100.4Hz in Fig.\ref{fig3}(b). We choose the appropriate oscillator resonance frequency scan step $\Delta \omega_0$ from this.

According to the calculation results from Fig.\ref{fig3}(a) and
Fig.\ref{fig3}(b), we choose the scan step $\Delta \omega_0=1$Hz in the region resonance frequency $ \omega_0 $ range from $0.1$Hz to $100$Hz, each scan cover the frequency range from  $\omega_0-\Delta \omega_0/2$ to $\omega_0+\Delta \omega_0/2$, and fix the laser input power $P_{\mathrm{in}}=P_{\mathrm{opt}}(\omega_0,\omega_0 )$ in each scan meanwhile. 
We calculate the acceleration measurement noise $S_{\mathrm{aa}}^{\mathrm{mea}}$ with $\eta=0.1$ in each scan, and calculate the envelope of these series $S_{\mathrm{aa}}^{\mathrm{mea}}$ writen as $S_{\mathrm{aa}}^{\mathrm{mea}^{\prime}}$. The acceleration measurement sensitivity $S_{\mathrm{aa}}^{\mathrm{tot}}=S_{\mathrm{aa}}^{\mathrm{th}}+S_{\mathrm{aa}}^{\mathrm{mea}^{\prime}}$, and these results are presented in Fig.\ref{fig3}(c).

\begin{figure}
	\centering
	\includegraphics[width=0.95\columnwidth]{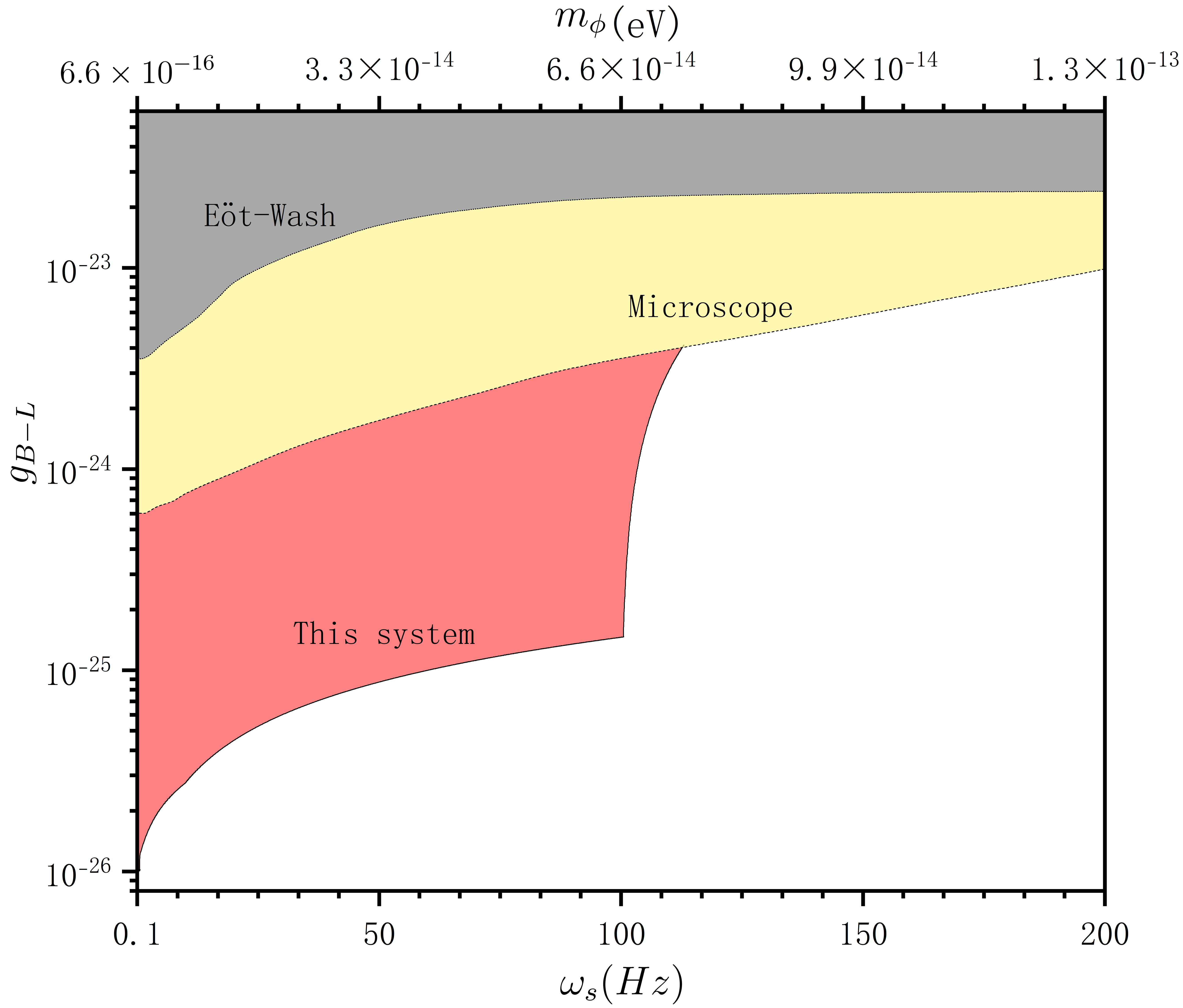}
	\caption{Ultra-light Dark Matter search range. The top axis represents the DM mass $m_{\phi}$ corresponding to the frequency $\omega_s$.
The upper grey and yellow regions are excluded by E$\mathrm{\ddot{o}}$t-Wash torsion balance \cite{Wagner_2012, PhysRevLett.100.041101, PhysRevLett.116.031102} and MICROSCOPE experiments \cite{PhysRevD.98.064051, PhysRevLett.120.141101}, and the red region is the range that this system can cover. In  torsion balance system, they use  a pair of accelerometers (Beryllium and Titanium) with a differential neutron/nucleon ratio $\Delta=\mathrm{N}_1/\mathrm{A}_1-\mathrm{N}_2/\mathrm{A}_2=0.037$, where N and A are the neutron and nucleon numbers of Beryllium and Titanium respectively. From the equation(\ref{Fsig}), $N_g$ can be approximated as $N_g=\Delta \cdot m/m_{\mathrm{neu}}$ at this time.}\label{fig4}
\end{figure}

According to the previous discussion on the effective integration time $T_{\mathrm{tot}}$,
we fix the measurement time of each scan as $T_{\mathrm{mea}}=10^5$s.
When DM frequency $\omega_s\textless10$Hz, $T_{\mathrm{tot}}=T_{\mathrm{mea}}$; and when $\omega_s\textgreater10$Hz, $T_{\mathrm{tot}}=\sqrt{T_{\mathrm{mea}} \cdot 10^6/\omega_s}$.
Combining previous discussion of the scan step, we estimate that  about one  hundred times adjustments and measurements will be required in total,  corresponding to a total time of $1 \times 10^7$ seconds.
The final result of coupling strength $g_{\scriptscriptstyle B-L}$ from equation (\ref{gBL}) is shown in Fig.\ref{fig4}.  In the region of $\omega_s \textless 100$Hz, this system always has high acceleration sensitivity by adjusting the resonance frequency of the mechanical oscillator. And we achieve more than an order of magnitude improvement in the measurement of $g_{\scriptscriptstyle B-L} $ compare to the MICROSCOPE and the Eöt-Wash torsion experiment. 
And in the region of $\omega_s \textgreater 100$Hz, the measurement accuracy of $g_{\scriptscriptstyle B-L} $ decreases rapidly, due to the increase in measurement noise $S_{\mathrm{aa}}^{\mathrm{mea}}$.

\begin{figure}
 	\centering
 	\includegraphics[width=0.95\columnwidth]{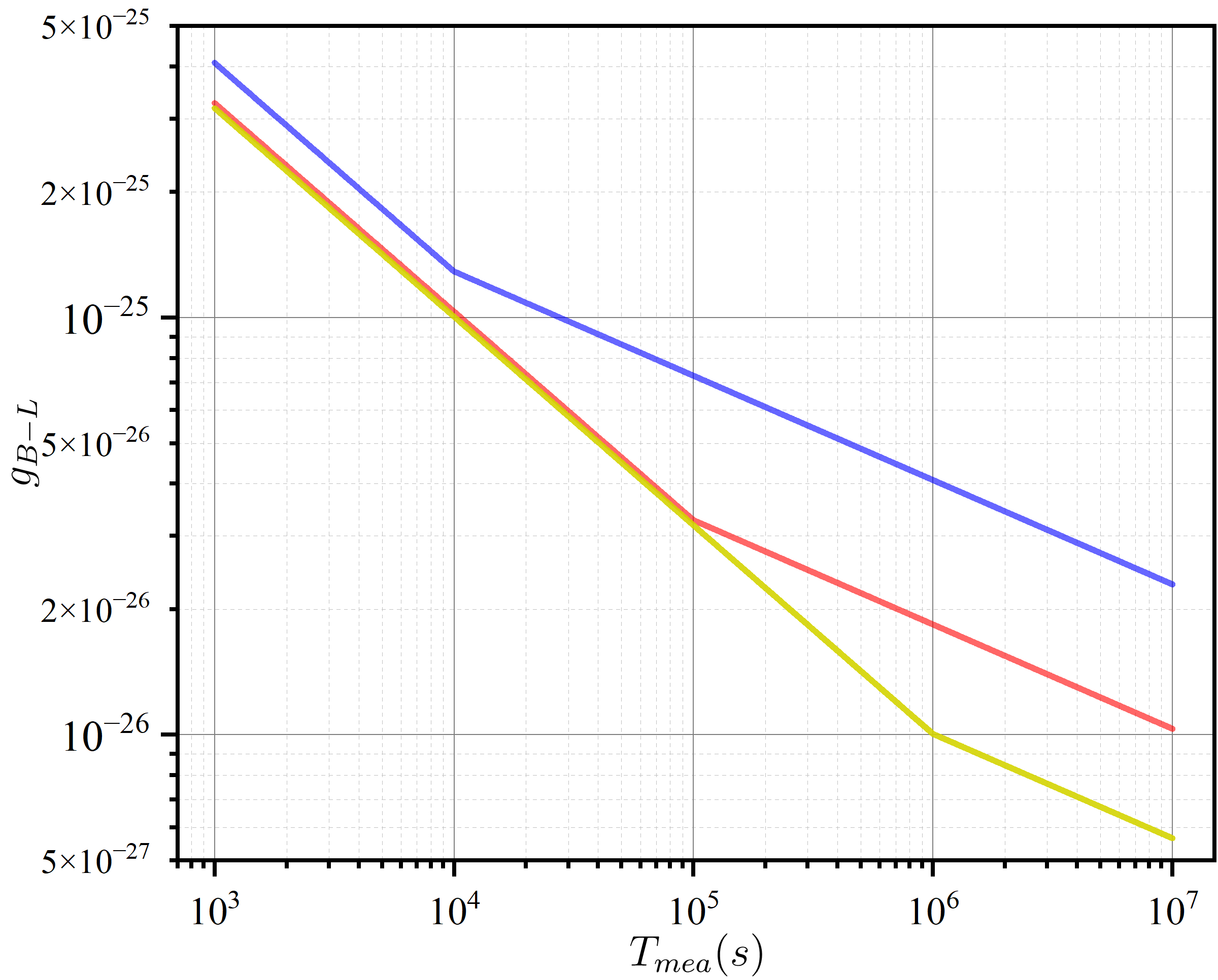}
 	\caption{The minimum $g_{\scriptscriptstyle B-L}$ can reach in different DM frequency $\omega_{s}$. The yellow, red and blue line  indicates $\omega_{s}$ is 1Hz, 10Hz and 100Hz respectively. The maximum measurement time is $10^7$s (about 115.7 days). }	\label{gBL-Tmea}
 \end{figure}

Finally, we estimated the minimum $g_{\scriptscriptstyle B-L}$ that this system can detect. Assume that the DM frequency $\omega_{s}$ is 1Hz, 10Hz and 100Hz respectively.
 From the equation (\ref{gBL}) and the measurement time $T_{\mathrm{mea}}$ range from $10^3$s to $10^7$s, the results are shown in Fig.\ref{gBL-Tmea}. 
 When $T_{\mathrm{mea}}$ is less than the coherent time $T_{\mathrm{coh}}$, $g_{\scriptscriptstyle B-L}$ decreases rapidly as $T_{\mathrm{mea}}$ increases; and when $T_{\mathrm{mea}}$ is greater than $T_{\mathrm{coh}}$, $g_{\scriptscriptstyle B-L}$ decreases more slowly. If the final measurement time is about $10^7 s$, the minimum $g_{\scriptscriptstyle B-L}$ that can be measured scale is about $10^{-26}$.

\section{CONCLUSION}
We propose an experimental scheme to detect ultra-light dark matter using a frequency adjustable diamagnetic levitated microsphere sensor which can theoretically approach the standard quantum limit. 
We change the resonance frequency by adjusting the distance between the  paramagnetic microsphere and the lower combined magnets, and  to obtain a lager range that maintains high acceleration measurement sensitivity.
Compared to the existing system, our method can achieve at least one order of magnitude improvement in the coupling constant $g_{\scriptscriptstyle B-L}$, especially in the frequencies from 0.1Hz to 100Hz. And it may be possible to achieve higher accuracy by using the array of sensors in the future.
 
 In this article, we consider only the effects of thermal noise and quantum measurement noise on the acceleration measurement sensitivity of the system. 
 In fact, there are many low frequency noises such as seismic waves and Earth tidal forces  which also have a great impact on the accuracy of the experiment, and that cannot be shielded by the suspension system. This poses a great challenge to the actual measurement. Reducing the frequency scan step according to the accuracy of the active vibration isolation device may make the effect of other noise lower than thermal noise, and this needs to be verified by further experiments.
  
In general, the current ground-based precision measurement system may have a broader prospect in terms of dark matter measurement compared to the previous astronomical observation methods. In the future, with the development of measurement sensitivity and 
measurement range of mechanical sensors , especially with the improvement quantum sensing technology, the measurement sensitivity may break through the standard quantum limit. It will open up more possibilities in terms of dark matter measurement.

\begin{acknowledgements}	
	This work was supported by the National Natural Science Foundation of China (Grants No.12205291, No. 12075115, No. 12075116, No. 11890702 and No. 12150011), the Fundamental Research	Funds for the Central Universities,  and Anhui Provincial Natural Science Foundation (Grant No. 2208085QA16).	
\end{acknowledgements}

\bibliographystyle{apsrev4-1}
\bibliography{ultralightDM}

%merlin.mbs apsrev4-1.bst 2010-07-25 4.21a (PWD, AO, DPC) hacked
%Control: key (0)
%Control: author (72) initials jnrlst
%Control: editor formatted (1) identically to author
%Control: production of article title (-1) disabled
%Control: page (0) single
%Control: year (1) truncated
%Control: production of eprint (0) enabled
\begin{thebibliography}{36}%
\makeatletter
\providecommand \@ifxundefined [1]{%
 \@ifx{#1\undefined}
}%
\providecommand \@ifnum [1]{%
 \ifnum #1\expandafter \@firstoftwo
 \else \expandafter \@secondoftwo
 \fi
}%
\providecommand \@ifx [1]{%
 \ifx #1\expandafter \@firstoftwo
 \else \expandafter \@secondoftwo
 \fi
}%
\providecommand \natexlab [1]{#1}%
\providecommand \enquote  [1]{``#1''}%
\providecommand \bibnamefont  [1]{#1}%
\providecommand \bibfnamefont [1]{#1}%
\providecommand \citenamefont [1]{#1}%
\providecommand \href@noop [0]{\@secondoftwo}%
\providecommand \href [0]{\begingroup \@sanitize@url \@href}%
\providecommand \@href[1]{\@@startlink{#1}\@@href}%
\providecommand \@@href[1]{\endgroup#1\@@endlink}%
\providecommand \@sanitize@url [0]{\catcode `\\12\catcode `\$12\catcode
  `\&12\catcode `\#12\catcode `\^12\catcode `\_12\catcode `\%12\relax}%
\providecommand \@@startlink[1]{}%
\providecommand \@@endlink[0]{}%
\providecommand \url  [0]{\begingroup\@sanitize@url \@url }%
\providecommand \@url [1]{\endgroup\@href {#1}{\urlprefix }}%
\providecommand \urlprefix  [0]{URL }%
\providecommand \Eprint [0]{\href }%
\providecommand \doibase [0]{http://dx.doi.org/}%
\providecommand \selectlanguage [0]{\@gobble}%
\providecommand \bibinfo  [0]{\@secondoftwo}%
\providecommand \bibfield  [0]{\@secondoftwo}%
\providecommand \translation [1]{[#1]}%
\providecommand \BibitemOpen [0]{}%
\providecommand \bibitemStop [0]{}%
\providecommand \bibitemNoStop [0]{.\EOS\space}%
\providecommand \EOS [0]{\spacefactor3000\relax}%
\providecommand \BibitemShut  [1]{\csname bibitem#1\endcsname}%
\let\auto@bib@innerbib\@empty
%</preamble>
\bibitem [{\citenamefont {Sofue}\ and\ \citenamefont
  {Rubin}(2001)}]{doi:10.1146/annurev.astro.39.1.137}%
  \BibitemOpen
  \bibfield  {author} {\bibinfo {author} {\bibfnamefont {Y.}~\bibnamefont
  {Sofue}}\ and\ \bibinfo {author} {\bibfnamefont {V.}~\bibnamefont {Rubin}},\
  }\href {\doibase 10.1146/annurev.astro.39.1.137} {\bibfield  {journal}
  {\bibinfo  {journal} {Annual Review of Astronomy and Astrophysics}\ }\textbf
  {\bibinfo {volume} {39}},\ \bibinfo {pages} {137} (\bibinfo {year}
  {2001})}\BibitemShut {NoStop}%
\bibitem [{\citenamefont {Massey}\ \emph {et~al.}(2010)\citenamefont {Massey},
  \citenamefont {Kitching},\ and\ \citenamefont {Richard}}]{Massey_2010}%
  \BibitemOpen
  \bibfield  {author} {\bibinfo {author} {\bibfnamefont {R.}~\bibnamefont
  {Massey}}, \bibinfo {author} {\bibfnamefont {T.}~\bibnamefont {Kitching}}, \
  and\ \bibinfo {author} {\bibfnamefont {J.}~\bibnamefont {Richard}},\ }\href
  {\doibase 10.1088/0034-4885/73/8/086901} {\bibfield  {journal} {\bibinfo
  {journal} {Reports on Progress in Physics}\ }\textbf {\bibinfo {volume}
  {73}},\ \bibinfo {pages} {086901} (\bibinfo {year} {2010})}\BibitemShut
  {NoStop}%
\bibitem [{\citenamefont {Markevitch}\ \emph {et~al.}(2004)\citenamefont
  {Markevitch}, \citenamefont {Gonzalez}, \citenamefont {Clowe}, \citenamefont
  {Vikhlinin}, \citenamefont {Forman}, \citenamefont {Jones}, \citenamefont
  {Murray},\ and\ \citenamefont {Tucker}}]{Markevitch_2004}%
  \BibitemOpen
  \bibfield  {author} {\bibinfo {author} {\bibfnamefont {M.}~\bibnamefont
  {Markevitch}}, \bibinfo {author} {\bibfnamefont {A.~H.}\ \bibnamefont
  {Gonzalez}}, \bibinfo {author} {\bibfnamefont {D.}~\bibnamefont {Clowe}},
  \bibinfo {author} {\bibfnamefont {A.}~\bibnamefont {Vikhlinin}}, \bibinfo
  {author} {\bibfnamefont {W.}~\bibnamefont {Forman}}, \bibinfo {author}
  {\bibfnamefont {C.}~\bibnamefont {Jones}}, \bibinfo {author} {\bibfnamefont
  {S.}~\bibnamefont {Murray}}, \ and\ \bibinfo {author} {\bibfnamefont
  {W.}~\bibnamefont {Tucker}},\ }\href {\doibase 10.1086/383178} {\bibfield
  {journal} {\bibinfo  {journal} {The Astrophysical Journal}\ }\textbf
  {\bibinfo {volume} {606}},\ \bibinfo {pages} {819} (\bibinfo {year}
  {2004})}\BibitemShut {NoStop}%
\bibitem [{\citenamefont {Gianfranco}(2010)}]{bertone_2010}%
  \BibitemOpen
  \bibfield  {author} {\bibinfo {author} {\bibfnamefont {B.}~\bibnamefont
  {Gianfranco}},\ }\href {\doibase 10.1017/CBO9780511770739} {\emph {\bibinfo
  {title} {Particle Dark Matter: Observations, Models and Searches}}}\
  (\bibinfo  {publisher} {Cambridge University Press},\ \bibinfo {year}
  {2010})\BibitemShut {NoStop}%
\bibitem [{\citenamefont {Salucci}(2019)}]{Salucci_2019}%
  \BibitemOpen
  \bibfield  {author} {\bibinfo {author} {\bibfnamefont {P.}~\bibnamefont
  {Salucci}},\ }\href {\doibase 10.1007/s00159-018-0113-1} {\bibfield
  {journal} {\bibinfo  {journal} {The Astronomy and Astrophysics Review}\
  }\textbf {\bibinfo {volume} {27}},\ \bibinfo {pages} {2} (\bibinfo {year}
  {2019})}\BibitemShut {NoStop}%
\bibitem [{\citenamefont {Tanabashi}\ \emph {et~al.}(2018)\citenamefont
  {Tanabashi}, \citenamefont {Hagiwara}, \citenamefont {Hikasa}, \citenamefont
  {Nakamura} \emph {et~al.}}]{PhysRevD.98.030001}%
  \BibitemOpen
  \bibfield  {author} {\bibinfo {author} {\bibfnamefont {M.}~\bibnamefont
  {Tanabashi}}, \bibinfo {author} {\bibfnamefont {K.}~\bibnamefont {Hagiwara}},
  \bibinfo {author} {\bibfnamefont {K.}~\bibnamefont {Hikasa}}, \bibinfo
  {author} {\bibfnamefont {K.}~\bibnamefont {Nakamura}},  \emph {et~al.}
  (\bibinfo {collaboration} {Particle Data Group}),\ }\href {\doibase
  10.1103/PhysRevD.98.030001} {\bibfield  {journal} {\bibinfo  {journal} {Phys.
  Rev. D}\ }\textbf {\bibinfo {volume} {98}},\ \bibinfo {pages} {030001}
  (\bibinfo {year} {2018})}\BibitemShut {NoStop}%
\bibitem [{\citenamefont {Irastorza}\ and\ \citenamefont
  {Redondo}(2018)}]{IRASTORZA201889}%
  \BibitemOpen
  \bibfield  {author} {\bibinfo {author} {\bibfnamefont {I.~G.}\ \bibnamefont
  {Irastorza}}\ and\ \bibinfo {author} {\bibfnamefont {J.}~\bibnamefont
  {Redondo}},\ }\href {\doibase https://doi.org/10.1016/j.ppnp.2018.05.003}
  {\bibfield  {journal} {\bibinfo  {journal} {Progress in Particle and Nuclear
  Physics}\ }\textbf {\bibinfo {volume} {102}},\ \bibinfo {pages} {89}
  (\bibinfo {year} {2018})}\BibitemShut {NoStop}%
\bibitem [{\citenamefont {Hui}\ \emph {et~al.}(2017)\citenamefont {Hui},
  \citenamefont {Ostriker}, \citenamefont {Tremaine},\ and\ \citenamefont
  {Witten}}]{PhysRevD.95.043541}%
  \BibitemOpen
  \bibfield  {author} {\bibinfo {author} {\bibfnamefont {L.}~\bibnamefont
  {Hui}}, \bibinfo {author} {\bibfnamefont {J.~P.}\ \bibnamefont {Ostriker}},
  \bibinfo {author} {\bibfnamefont {S.}~\bibnamefont {Tremaine}}, \ and\
  \bibinfo {author} {\bibfnamefont {E.}~\bibnamefont {Witten}},\ }\href
  {\doibase 10.1103/PhysRevD.95.043541} {\bibfield  {journal} {\bibinfo
  {journal} {Phys. Rev. D}\ }\textbf {\bibinfo {volume} {95}},\ \bibinfo
  {pages} {043541} (\bibinfo {year} {2017})}\BibitemShut {NoStop}%
\bibitem [{\citenamefont {Bovy}\ \emph {et~al.}(2012)\citenamefont {Bovy},
  \citenamefont {Prieto}, \citenamefont {Beers}, \citenamefont {Bizyaev} \emph
  {et~al.}}]{Bovy_2012}%
  \BibitemOpen
  \bibfield  {author} {\bibinfo {author} {\bibfnamefont {J.}~\bibnamefont
  {Bovy}}, \bibinfo {author} {\bibfnamefont {C.~A.}\ \bibnamefont {Prieto}},
  \bibinfo {author} {\bibfnamefont {T.~C.}\ \bibnamefont {Beers}}, \bibinfo
  {author} {\bibfnamefont {D.}~\bibnamefont {Bizyaev}},  \emph {et~al.},\
  }\href {\doibase 10.1088/0004-637X/759/2/131} {\bibfield  {journal} {\bibinfo
   {journal} {The Astrophysical Journal}\ }\textbf {\bibinfo {volume} {759}},\
  \bibinfo {pages} {131} (\bibinfo {year} {2012})}\BibitemShut {NoStop}%
\bibitem [{\citenamefont {O'Hare}\ \emph {et~al.}(2018)\citenamefont {O'Hare},
  \citenamefont {McCabe}, \citenamefont {Evans}, \citenamefont {Myeong},\ and\
  \citenamefont {Belokurov}}]{PhysRevD.98.103006}%
  \BibitemOpen
  \bibfield  {author} {\bibinfo {author} {\bibfnamefont {C.~A.~J.}\
  \bibnamefont {O'Hare}}, \bibinfo {author} {\bibfnamefont {C.}~\bibnamefont
  {McCabe}}, \bibinfo {author} {\bibfnamefont {N.~W.}\ \bibnamefont {Evans}},
  \bibinfo {author} {\bibfnamefont {G.}~\bibnamefont {Myeong}}, \ and\ \bibinfo
  {author} {\bibfnamefont {V.}~\bibnamefont {Belokurov}},\ }\href {\doibase
  10.1103/PhysRevD.98.103006} {\bibfield  {journal} {\bibinfo  {journal} {Phys.
  Rev. D}\ }\textbf {\bibinfo {volume} {98}},\ \bibinfo {pages} {103006}
  (\bibinfo {year} {2018})}\BibitemShut {NoStop}%
\bibitem [{\citenamefont {Myeong}\ \emph {et~al.}(2017)\citenamefont {Myeong},
  \citenamefont {Evans}, \citenamefont {Belokurov}, \citenamefont {Amorisco},\
  and\ \citenamefont {Koposov}}]{10.1093/mnras/stx3262}%
  \BibitemOpen
  \bibfield  {author} {\bibinfo {author} {\bibfnamefont {G.~C.}\ \bibnamefont
  {Myeong}}, \bibinfo {author} {\bibfnamefont {N.~W.}\ \bibnamefont {Evans}},
  \bibinfo {author} {\bibfnamefont {V.}~\bibnamefont {Belokurov}}, \bibinfo
  {author} {\bibfnamefont {N.~C.}\ \bibnamefont {Amorisco}}, \ and\ \bibinfo
  {author} {\bibfnamefont {S.~E.}\ \bibnamefont {Koposov}},\ }\href {\doibase
  10.1093/mnras/stx3262} {\bibfield  {journal} {\bibinfo  {journal} {Monthly
  Notices of the Royal Astronomical Society}\ }\textbf {\bibinfo {volume}
  {475}},\ \bibinfo {pages} {1537} (\bibinfo {year} {2017})}\BibitemShut
  {NoStop}%
\bibitem [{\citenamefont {Du}\ \emph {et~al.}(2018)\citenamefont {Du},
  \citenamefont {Force}, \citenamefont {Khatiwada}, \citenamefont {Lentz} \emph
  {et~al.}}]{PhysRevLett.120.151301}%
  \BibitemOpen
  \bibfield  {author} {\bibinfo {author} {\bibfnamefont {N.}~\bibnamefont
  {Du}}, \bibinfo {author} {\bibfnamefont {N.}~\bibnamefont {Force}}, \bibinfo
  {author} {\bibfnamefont {R.}~\bibnamefont {Khatiwada}}, \bibinfo {author}
  {\bibfnamefont {E.}~\bibnamefont {Lentz}},  \emph {et~al.} (\bibinfo
  {collaboration} {ADMX Collaboration}),\ }\href {\doibase
  10.1103/PhysRevLett.120.151301} {\bibfield  {journal} {\bibinfo  {journal}
  {Phys. Rev. Lett.}\ }\textbf {\bibinfo {volume} {120}},\ \bibinfo {pages}
  {151301} (\bibinfo {year} {2018})}\BibitemShut {NoStop}%
\bibitem [{\citenamefont {Backes}\ \emph {et~al.}(2021)\citenamefont {Backes},
  \citenamefont {Palken}, \citenamefont {Kenany} \emph
  {et~al.}}]{backes2021quantum}%
  \BibitemOpen
  \bibfield  {author} {\bibinfo {author} {\bibfnamefont {K.~M.}\ \bibnamefont
  {Backes}}, \bibinfo {author} {\bibfnamefont {D.~A.}\ \bibnamefont {Palken}},
  \bibinfo {author} {\bibfnamefont {S.~A.}\ \bibnamefont {Kenany}},  \emph
  {et~al.},\ }\href {\doibase 10.1038/s41586-021-03226-7} {\bibfield  {journal}
  {\bibinfo  {journal} {Nature}\ }\textbf {\bibinfo {volume} {590}},\ \bibinfo
  {pages} {238} (\bibinfo {year} {2021})}\BibitemShut {NoStop}%
\bibitem [{\citenamefont {Kwon}\ \emph {et~al.}(2021)\citenamefont {Kwon},
  \citenamefont {Lee}, \citenamefont {Chung}, \citenamefont {Ahn},
  \citenamefont {Byun} \emph {et~al.}}]{PhysRevLett.126.191802}%
  \BibitemOpen
  \bibfield  {author} {\bibinfo {author} {\bibfnamefont {O.}~\bibnamefont
  {Kwon}}, \bibinfo {author} {\bibfnamefont {D.}~\bibnamefont {Lee}}, \bibinfo
  {author} {\bibfnamefont {W.}~\bibnamefont {Chung}}, \bibinfo {author}
  {\bibfnamefont {D.}~\bibnamefont {Ahn}}, \bibinfo {author} {\bibfnamefont
  {H.}~\bibnamefont {Byun}},  \emph {et~al.},\ }\href {\doibase
  10.1103/PhysRevLett.126.191802} {\bibfield  {journal} {\bibinfo  {journal}
  {Phys. Rev. Lett.}\ }\textbf {\bibinfo {volume} {126}},\ \bibinfo {pages}
  {191802} (\bibinfo {year} {2021})}\BibitemShut {NoStop}%
\bibitem [{\citenamefont {Abel}\ \emph {et~al.}(2017)\citenamefont {Abel},
  \citenamefont {Ayres}, \citenamefont {Ban}, \citenamefont {Bison} \emph
  {et~al.}}]{PhysRevX.7.041034}%
  \BibitemOpen
  \bibfield  {author} {\bibinfo {author} {\bibfnamefont {C.}~\bibnamefont
  {Abel}}, \bibinfo {author} {\bibfnamefont {N.~J.}\ \bibnamefont {Ayres}},
  \bibinfo {author} {\bibfnamefont {G.}~\bibnamefont {Ban}}, \bibinfo {author}
  {\bibfnamefont {G.}~\bibnamefont {Bison}},  \emph {et~al.},\ }\href {\doibase
  10.1103/PhysRevX.7.041034} {\bibfield  {journal} {\bibinfo  {journal} {Phys.
  Rev. X}\ }\textbf {\bibinfo {volume} {7}},\ \bibinfo {pages} {041034}
  (\bibinfo {year} {2017})}\BibitemShut {NoStop}%
\bibitem [{\citenamefont {Terrano}\ \emph {et~al.}(2019)\citenamefont
  {Terrano}, \citenamefont {Adelberger}, \citenamefont {Hagedorn},\ and\
  \citenamefont {Heckel}}]{PhysRevLett.122.231301}%
  \BibitemOpen
  \bibfield  {author} {\bibinfo {author} {\bibfnamefont {W.~A.}\ \bibnamefont
  {Terrano}}, \bibinfo {author} {\bibfnamefont {E.~G.}\ \bibnamefont
  {Adelberger}}, \bibinfo {author} {\bibfnamefont {C.~A.}\ \bibnamefont
  {Hagedorn}}, \ and\ \bibinfo {author} {\bibfnamefont {B.~R.}\ \bibnamefont
  {Heckel}},\ }\href {\doibase 10.1103/PhysRevLett.122.231301} {\bibfield
  {journal} {\bibinfo  {journal} {Phys. Rev. Lett.}\ }\textbf {\bibinfo
  {volume} {122}},\ \bibinfo {pages} {231301} (\bibinfo {year}
  {2019})}\BibitemShut {NoStop}%
\bibitem [{\citenamefont {Smorra}\ \emph {et~al.}(2019)\citenamefont {Smorra},
  \citenamefont {Stadnik}, \citenamefont {Blessing}, \citenamefont {Bohman},
  \citenamefont {Borchert}, \citenamefont {Devlin}, \citenamefont {Erlewein},
  \citenamefont {Harrington}, \citenamefont {Higuchi}, \citenamefont {Mooser}
  \emph {et~al.}}]{smorra2019direct}%
  \BibitemOpen
  \bibfield  {author} {\bibinfo {author} {\bibfnamefont {C.}~\bibnamefont
  {Smorra}}, \bibinfo {author} {\bibfnamefont {Y.}~\bibnamefont {Stadnik}},
  \bibinfo {author} {\bibfnamefont {P.}~\bibnamefont {Blessing}}, \bibinfo
  {author} {\bibfnamefont {M.}~\bibnamefont {Bohman}}, \bibinfo {author}
  {\bibfnamefont {M.}~\bibnamefont {Borchert}}, \bibinfo {author}
  {\bibfnamefont {J.}~\bibnamefont {Devlin}}, \bibinfo {author} {\bibfnamefont
  {S.}~\bibnamefont {Erlewein}}, \bibinfo {author} {\bibfnamefont
  {J.}~\bibnamefont {Harrington}}, \bibinfo {author} {\bibfnamefont
  {T.}~\bibnamefont {Higuchi}}, \bibinfo {author} {\bibfnamefont
  {A.}~\bibnamefont {Mooser}},  \emph {et~al.},\ }\href {\doibase
  10.1038/s41586-019-1727-9} {\bibfield  {journal} {\bibinfo  {journal}
  {Nature}\ }\textbf {\bibinfo {volume} {575}},\ \bibinfo {pages} {310}
  (\bibinfo {year} {2019})}\BibitemShut {NoStop}%
\bibitem [{\citenamefont {Kennedy}\ \emph {et~al.}(2020)\citenamefont
  {Kennedy}, \citenamefont {Oelker}, \citenamefont {Robinson}, \citenamefont
  {Bothwell}, \citenamefont {Kedar}, \citenamefont {Milner}, \citenamefont
  {Marti}, \citenamefont {Derevianko},\ and\ \citenamefont
  {Ye}}]{PhysRevLett.125.201302}%
  \BibitemOpen
  \bibfield  {author} {\bibinfo {author} {\bibfnamefont {C.~J.}\ \bibnamefont
  {Kennedy}}, \bibinfo {author} {\bibfnamefont {E.}~\bibnamefont {Oelker}},
  \bibinfo {author} {\bibfnamefont {J.~M.}\ \bibnamefont {Robinson}}, \bibinfo
  {author} {\bibfnamefont {T.}~\bibnamefont {Bothwell}}, \bibinfo {author}
  {\bibfnamefont {D.}~\bibnamefont {Kedar}}, \bibinfo {author} {\bibfnamefont
  {W.~R.}\ \bibnamefont {Milner}}, \bibinfo {author} {\bibfnamefont {G.~E.}\
  \bibnamefont {Marti}}, \bibinfo {author} {\bibfnamefont {A.}~\bibnamefont
  {Derevianko}}, \ and\ \bibinfo {author} {\bibfnamefont {J.}~\bibnamefont
  {Ye}},\ }\href {\doibase 10.1103/PhysRevLett.125.201302} {\bibfield
  {journal} {\bibinfo  {journal} {Phys. Rev. Lett.}\ }\textbf {\bibinfo
  {volume} {125}},\ \bibinfo {pages} {201302} (\bibinfo {year}
  {2020})}\BibitemShut {NoStop}%
\bibitem [{\citenamefont {Arvanitaki}\ \emph {et~al.}(2015)\citenamefont
  {Arvanitaki}, \citenamefont {Huang},\ and\ \citenamefont
  {Van~Tilburg}}]{PhysRevD.91.015015}%
  \BibitemOpen
  \bibfield  {author} {\bibinfo {author} {\bibfnamefont {A.}~\bibnamefont
  {Arvanitaki}}, \bibinfo {author} {\bibfnamefont {J.}~\bibnamefont {Huang}}, \
  and\ \bibinfo {author} {\bibfnamefont {K.}~\bibnamefont {Van~Tilburg}},\
  }\href {\doibase 10.1103/PhysRevD.91.015015} {\bibfield  {journal} {\bibinfo
  {journal} {Phys. Rev. D}\ }\textbf {\bibinfo {volume} {91}},\ \bibinfo
  {pages} {015015} (\bibinfo {year} {2015})}\BibitemShut {NoStop}%
\bibitem [{\citenamefont {Vermeulen}\ \emph {et~al.}(2021)\citenamefont
  {Vermeulen}, \citenamefont {Relton}, \citenamefont {Grote}, \citenamefont
  {Raymond}, \citenamefont {Affeldt}, \citenamefont {Bergamin}, \citenamefont
  {Bisht}, \citenamefont {Brinkmann}, \citenamefont {Danzmann}, \citenamefont
  {Doravari} \emph {et~al.}}]{vermeulen2021direct}%
  \BibitemOpen
  \bibfield  {author} {\bibinfo {author} {\bibfnamefont {S.~M.}\ \bibnamefont
  {Vermeulen}}, \bibinfo {author} {\bibfnamefont {P.}~\bibnamefont {Relton}},
  \bibinfo {author} {\bibfnamefont {H.}~\bibnamefont {Grote}}, \bibinfo
  {author} {\bibfnamefont {V.}~\bibnamefont {Raymond}}, \bibinfo {author}
  {\bibfnamefont {C.}~\bibnamefont {Affeldt}}, \bibinfo {author} {\bibfnamefont
  {F.}~\bibnamefont {Bergamin}}, \bibinfo {author} {\bibfnamefont
  {A.}~\bibnamefont {Bisht}}, \bibinfo {author} {\bibfnamefont
  {M.}~\bibnamefont {Brinkmann}}, \bibinfo {author} {\bibfnamefont
  {K.}~\bibnamefont {Danzmann}}, \bibinfo {author} {\bibfnamefont
  {S.}~\bibnamefont {Doravari}},  \emph {et~al.},\ }\href {\doibase
  10.1038/s41586-021-04031-y} {\bibfield  {journal} {\bibinfo  {journal}
  {Nature}\ }\textbf {\bibinfo {volume} {600}},\ \bibinfo {pages} {424}
  (\bibinfo {year} {2021})}\BibitemShut {NoStop}%
\bibitem [{\citenamefont {Carney}\ \emph {et~al.}(2021)\citenamefont {Carney},
  \citenamefont {Hook}, \citenamefont {Liu}, \citenamefont {Taylor},\ and\
  \citenamefont {Zhao}}]{Carney_2021}%
  \BibitemOpen
  \bibfield  {author} {\bibinfo {author} {\bibfnamefont {D.}~\bibnamefont
  {Carney}}, \bibinfo {author} {\bibfnamefont {A.}~\bibnamefont {Hook}},
  \bibinfo {author} {\bibfnamefont {Z.}~\bibnamefont {Liu}}, \bibinfo {author}
  {\bibfnamefont {J.~M.}\ \bibnamefont {Taylor}}, \ and\ \bibinfo {author}
  {\bibfnamefont {Y.}~\bibnamefont {Zhao}},\ }\href {\doibase
  10.1088/1367-2630/abd9e7} {\bibfield  {journal} {\bibinfo  {journal} {New
  Journal of Physics}\ }\textbf {\bibinfo {volume} {23}},\ \bibinfo {pages}
  {023041} (\bibinfo {year} {2021})}\BibitemShut {NoStop}%
\bibitem [{\citenamefont {Read}(2014)}]{Read_2014}%
  \BibitemOpen
  \bibfield  {author} {\bibinfo {author} {\bibfnamefont {J.~I.}\ \bibnamefont
  {Read}},\ }\href {\doibase 10.1088/0954-3899/41/6/063101} {\bibfield
  {journal} {\bibinfo  {journal} {Journal of Physics G: Nuclear and Particle
  Physics}\ }\textbf {\bibinfo {volume} {41}},\ \bibinfo {pages} {063101}
  (\bibinfo {year} {2014})}\BibitemShut {NoStop}%
\bibitem [{\citenamefont {Clark}\ \emph {et~al.}(2016)\citenamefont {Clark},
  \citenamefont {Lecocq}, \citenamefont {Simmonds}, \citenamefont {Aumentado},\
  and\ \citenamefont {Teufel}}]{Clark2016SidebandCB}%
  \BibitemOpen
  \bibfield  {author} {\bibinfo {author} {\bibfnamefont {J.~B.}\ \bibnamefont
  {Clark}}, \bibinfo {author} {\bibfnamefont {F.}~\bibnamefont {Lecocq}},
  \bibinfo {author} {\bibfnamefont {R.~W.}\ \bibnamefont {Simmonds}}, \bibinfo
  {author} {\bibfnamefont {J.}~\bibnamefont {Aumentado}}, \ and\ \bibinfo
  {author} {\bibfnamefont {J.~D.}\ \bibnamefont {Teufel}},\ }\href {\doibase
  10.1103/RevModPhys.82.1155} {\bibfield  {journal} {\bibinfo  {journal}
  {Nature}\ }\textbf {\bibinfo {volume} {541}},\ \bibinfo {pages} {191}
  (\bibinfo {year} {2016})}\BibitemShut {NoStop}%
\bibitem [{\citenamefont {Leng}\ \emph {et~al.}(2021)\citenamefont {Leng},
  \citenamefont {Li}, \citenamefont {Kong}, \citenamefont {Xie}, \citenamefont
  {Zheng}, \citenamefont {Yin}, \citenamefont {Xiong}, \citenamefont {Wu},
  \citenamefont {Duan}, \citenamefont {Du}, \citenamefont {Yin}, \citenamefont
  {Huang},\ and\ \citenamefont {Du}}]{PhysRevApplied.15.024061}%
  \BibitemOpen
  \bibfield  {author} {\bibinfo {author} {\bibfnamefont {Y.}~\bibnamefont
  {Leng}}, \bibinfo {author} {\bibfnamefont {R.}~\bibnamefont {Li}}, \bibinfo
  {author} {\bibfnamefont {X.}~\bibnamefont {Kong}}, \bibinfo {author}
  {\bibfnamefont {H.}~\bibnamefont {Xie}}, \bibinfo {author} {\bibfnamefont
  {D.}~\bibnamefont {Zheng}}, \bibinfo {author} {\bibfnamefont
  {P.}~\bibnamefont {Yin}}, \bibinfo {author} {\bibfnamefont {F.}~\bibnamefont
  {Xiong}}, \bibinfo {author} {\bibfnamefont {T.}~\bibnamefont {Wu}}, \bibinfo
  {author} {\bibfnamefont {C.-K.}\ \bibnamefont {Duan}}, \bibinfo {author}
  {\bibfnamefont {Y.}~\bibnamefont {Du}}, \bibinfo {author} {\bibfnamefont
  {Z.-q.}\ \bibnamefont {Yin}}, \bibinfo {author} {\bibfnamefont
  {P.}~\bibnamefont {Huang}}, \ and\ \bibinfo {author} {\bibfnamefont
  {J.}~\bibnamefont {Du}},\ }\href {\doibase 10.1103/PhysRevApplied.15.024061}
  {\bibfield  {journal} {\bibinfo  {journal} {Phys. Rev. Appl.}\ }\textbf
  {\bibinfo {volume} {15}},\ \bibinfo {pages} {024061} (\bibinfo {year}
  {2021})}\BibitemShut {NoStop}%
\bibitem [{\citenamefont {Monteiro}\ \emph {et~al.}(2020)\citenamefont
  {Monteiro}, \citenamefont {Li}, \citenamefont {Afek}, \citenamefont {Li},
  \citenamefont {Mossman},\ and\ \citenamefont {Moore}}]{PhysRevA.101.053835}%
  \BibitemOpen
  \bibfield  {author} {\bibinfo {author} {\bibfnamefont {F.}~\bibnamefont
  {Monteiro}}, \bibinfo {author} {\bibfnamefont {W.}~\bibnamefont {Li}},
  \bibinfo {author} {\bibfnamefont {G.}~\bibnamefont {Afek}}, \bibinfo {author}
  {\bibfnamefont {C.-l.}\ \bibnamefont {Li}}, \bibinfo {author} {\bibfnamefont
  {M.}~\bibnamefont {Mossman}}, \ and\ \bibinfo {author} {\bibfnamefont
  {D.~C.}\ \bibnamefont {Moore}},\ }\href {\doibase
  10.1103/PhysRevA.101.053835} {\bibfield  {journal} {\bibinfo  {journal}
  {Phys. Rev. A}\ }\textbf {\bibinfo {volume} {101}},\ \bibinfo {pages}
  {053835} (\bibinfo {year} {2020})}\BibitemShut {NoStop}%
\bibitem [{\citenamefont {Acernese}\ \emph {et~al.}(2010)\citenamefont
  {Acernese}, \citenamefont {Antonucci}, \citenamefont {Aoudia}, \citenamefont
  {Arun}, \citenamefont {Astone}, \citenamefont {Ballardin} \emph
  {et~al.}}]{ACERNESE2010182}%
  \BibitemOpen
  \bibfield  {author} {\bibinfo {author} {\bibfnamefont {F.}~\bibnamefont
  {Acernese}}, \bibinfo {author} {\bibfnamefont {F.}~\bibnamefont {Antonucci}},
  \bibinfo {author} {\bibfnamefont {S.}~\bibnamefont {Aoudia}}, \bibinfo
  {author} {\bibfnamefont {K.}~\bibnamefont {Arun}}, \bibinfo {author}
  {\bibfnamefont {P.}~\bibnamefont {Astone}}, \bibinfo {author} {\bibfnamefont
  {G.}~\bibnamefont {Ballardin}},  \emph {et~al.},\ }\href {\doibase
  https://doi.org/10.1016/j.astropartphys.2010.01.006} {\bibfield  {journal}
  {\bibinfo  {journal} {Astroparticle Physics}\ }\textbf {\bibinfo {volume}
  {33}},\ \bibinfo {pages} {182} (\bibinfo {year} {2010})}\BibitemShut
  {NoStop}%
\bibitem [{\citenamefont {Acernese}\ \emph {et~al.}(2014)\citenamefont
  {Acernese}, \citenamefont {Agathos}, \citenamefont {Agatsuma}, \citenamefont
  {Aisa}, \citenamefont {Allemandou} \emph {et~al.}}]{Acernese_2015}%
  \BibitemOpen
  \bibfield  {author} {\bibinfo {author} {\bibfnamefont {F.}~\bibnamefont
  {Acernese}}, \bibinfo {author} {\bibfnamefont {M.}~\bibnamefont {Agathos}},
  \bibinfo {author} {\bibfnamefont {K.}~\bibnamefont {Agatsuma}}, \bibinfo
  {author} {\bibfnamefont {D.}~\bibnamefont {Aisa}}, \bibinfo {author}
  {\bibfnamefont {N.}~\bibnamefont {Allemandou}},  \emph {et~al.},\ }\href
  {\doibase 10.1088/0264-9381/32/2/024001} {\bibfield  {journal} {\bibinfo
  {journal} {Classical and Quantum Gravity}\ }\textbf {\bibinfo {volume}
  {32}},\ \bibinfo {pages} {024001} (\bibinfo {year} {2014})}\BibitemShut
  {NoStop}%
\bibitem [{\citenamefont {Zheng}\ \emph {et~al.}(2020)\citenamefont {Zheng},
  \citenamefont {Leng}, \citenamefont {Kong}, \citenamefont {Li}, \citenamefont
  {Wang}, \citenamefont {Luo}, \citenamefont {Zhao}, \citenamefont {Duan},
  \citenamefont {Huang}, \citenamefont {Du}, \citenamefont {Carlesso},\ and\
  \citenamefont {Bassi}}]{PhysRevResearch.2.013057}%
  \BibitemOpen
  \bibfield  {author} {\bibinfo {author} {\bibfnamefont {D.}~\bibnamefont
  {Zheng}}, \bibinfo {author} {\bibfnamefont {Y.}~\bibnamefont {Leng}},
  \bibinfo {author} {\bibfnamefont {X.}~\bibnamefont {Kong}}, \bibinfo {author}
  {\bibfnamefont {R.}~\bibnamefont {Li}}, \bibinfo {author} {\bibfnamefont
  {Z.}~\bibnamefont {Wang}}, \bibinfo {author} {\bibfnamefont {X.}~\bibnamefont
  {Luo}}, \bibinfo {author} {\bibfnamefont {J.}~\bibnamefont {Zhao}}, \bibinfo
  {author} {\bibfnamefont {C.-K.}\ \bibnamefont {Duan}}, \bibinfo {author}
  {\bibfnamefont {P.}~\bibnamefont {Huang}}, \bibinfo {author} {\bibfnamefont
  {J.}~\bibnamefont {Du}}, \bibinfo {author} {\bibfnamefont {M.}~\bibnamefont
  {Carlesso}}, \ and\ \bibinfo {author} {\bibfnamefont {A.}~\bibnamefont
  {Bassi}},\ }\href {\doibase 10.1103/PhysRevResearch.2.013057} {\bibfield
  {journal} {\bibinfo  {journal} {Phys. Rev. Res.}\ }\textbf {\bibinfo {volume}
  {2}},\ \bibinfo {pages} {013057} (\bibinfo {year} {2020})}\BibitemShut
  {NoStop}%
\bibitem [{\citenamefont {Xiong}\ \emph {et~al.}(2021)\citenamefont {Xiong},
  \citenamefont {Yin}, \citenamefont {Wu}, \citenamefont {Xie}, \citenamefont
  {Li}, \citenamefont {Leng}, \citenamefont {Li}, \citenamefont {Duan},
  \citenamefont {Kong}, \citenamefont {Huang},\ and\ \citenamefont
  {Du}}]{PhysRevApplied.16.L011003}%
  \BibitemOpen
  \bibfield  {author} {\bibinfo {author} {\bibfnamefont {F.}~\bibnamefont
  {Xiong}}, \bibinfo {author} {\bibfnamefont {P.}~\bibnamefont {Yin}}, \bibinfo
  {author} {\bibfnamefont {T.}~\bibnamefont {Wu}}, \bibinfo {author}
  {\bibfnamefont {H.}~\bibnamefont {Xie}}, \bibinfo {author} {\bibfnamefont
  {R.}~\bibnamefont {Li}}, \bibinfo {author} {\bibfnamefont {Y.}~\bibnamefont
  {Leng}}, \bibinfo {author} {\bibfnamefont {Y.}~\bibnamefont {Li}}, \bibinfo
  {author} {\bibfnamefont {C.}~\bibnamefont {Duan}}, \bibinfo {author}
  {\bibfnamefont {X.}~\bibnamefont {Kong}}, \bibinfo {author} {\bibfnamefont
  {P.}~\bibnamefont {Huang}}, \ and\ \bibinfo {author} {\bibfnamefont
  {J.}~\bibnamefont {Du}},\ }\href {\doibase 10.1103/PhysRevApplied.16.L011003}
  {\bibfield  {journal} {\bibinfo  {journal} {Phys. Rev. Appl.}\ }\textbf
  {\bibinfo {volume} {16}},\ \bibinfo {pages} {L011003} (\bibinfo {year}
  {2021})}\BibitemShut {NoStop}%
\bibitem [{\citenamefont {Wagner}\ \emph {et~al.}(2012)\citenamefont {Wagner},
  \citenamefont {Schlamminger}, \citenamefont {Gundlach},\ and\ \citenamefont
  {Adelberger}}]{Wagner_2012}%
  \BibitemOpen
  \bibfield  {author} {\bibinfo {author} {\bibfnamefont {T.~A.}\ \bibnamefont
  {Wagner}}, \bibinfo {author} {\bibfnamefont {S.}~\bibnamefont
  {Schlamminger}}, \bibinfo {author} {\bibfnamefont {J.~H.}\ \bibnamefont
  {Gundlach}}, \ and\ \bibinfo {author} {\bibfnamefont {E.~G.}\ \bibnamefont
  {Adelberger}},\ }\href {\doibase 10.1088/0264-9381/29/18/184002} {\bibfield
  {journal} {\bibinfo  {journal} {Classical and Quantum Gravity}\ }\textbf
  {\bibinfo {volume} {29}},\ \bibinfo {pages} {184002} (\bibinfo {year}
  {2012})}\BibitemShut {NoStop}%
\bibitem [{\citenamefont {Schlamminger}\ \emph {et~al.}(2008)\citenamefont
  {Schlamminger}, \citenamefont {Choi}, \citenamefont {Wagner}, \citenamefont
  {Gundlach},\ and\ \citenamefont {Adelberger}}]{PhysRevLett.100.041101}%
  \BibitemOpen
  \bibfield  {author} {\bibinfo {author} {\bibfnamefont {S.}~\bibnamefont
  {Schlamminger}}, \bibinfo {author} {\bibfnamefont {K.-Y.}\ \bibnamefont
  {Choi}}, \bibinfo {author} {\bibfnamefont {T.~A.}\ \bibnamefont {Wagner}},
  \bibinfo {author} {\bibfnamefont {J.~H.}\ \bibnamefont {Gundlach}}, \ and\
  \bibinfo {author} {\bibfnamefont {E.~G.}\ \bibnamefont {Adelberger}},\ }\href
  {\doibase 10.1103/PhysRevLett.100.041101} {\bibfield  {journal} {\bibinfo
  {journal} {Phys. Rev. Lett.}\ }\textbf {\bibinfo {volume} {100}},\ \bibinfo
  {pages} {041101} (\bibinfo {year} {2008})}\BibitemShut {NoStop}%
\bibitem [{\citenamefont {Arvanitaki}\ \emph {et~al.}(2016)\citenamefont
  {Arvanitaki}, \citenamefont {Dimopoulos},\ and\ \citenamefont
  {Van~Tilburg}}]{PhysRevLett.116.031102}%
  \BibitemOpen
  \bibfield  {author} {\bibinfo {author} {\bibfnamefont {A.}~\bibnamefont
  {Arvanitaki}}, \bibinfo {author} {\bibfnamefont {S.}~\bibnamefont
  {Dimopoulos}}, \ and\ \bibinfo {author} {\bibfnamefont {K.}~\bibnamefont
  {Van~Tilburg}},\ }\href {\doibase 10.1103/PhysRevLett.116.031102} {\bibfield
  {journal} {\bibinfo  {journal} {Phys. Rev. Lett.}\ }\textbf {\bibinfo
  {volume} {116}},\ \bibinfo {pages} {031102} (\bibinfo {year}
  {2016})}\BibitemShut {NoStop}%
\bibitem [{\citenamefont {Hees}\ \emph {et~al.}(2018)\citenamefont {Hees},
  \citenamefont {Minazzoli}, \citenamefont {Savalle}, \citenamefont {Stadnik},\
  and\ \citenamefont {Wolf}}]{PhysRevD.98.064051}%
  \BibitemOpen
  \bibfield  {author} {\bibinfo {author} {\bibfnamefont {A.}~\bibnamefont
  {Hees}}, \bibinfo {author} {\bibfnamefont {O.}~\bibnamefont {Minazzoli}},
  \bibinfo {author} {\bibfnamefont {E.}~\bibnamefont {Savalle}}, \bibinfo
  {author} {\bibfnamefont {Y.~V.}\ \bibnamefont {Stadnik}}, \ and\ \bibinfo
  {author} {\bibfnamefont {P.}~\bibnamefont {Wolf}},\ }\href {\doibase
  10.1103/PhysRevD.98.064051} {\bibfield  {journal} {\bibinfo  {journal} {Phys.
  Rev. D}\ }\textbf {\bibinfo {volume} {98}},\ \bibinfo {pages} {064051}
  (\bibinfo {year} {2018})}\BibitemShut {NoStop}%
\bibitem [{\citenamefont {Berg\'e}\ \emph {et~al.}(2018)\citenamefont
  {Berg\'e}, \citenamefont {Brax}, \citenamefont {M\'etris}, \citenamefont
  {Pernot-Borr\`as}, \citenamefont {Touboul},\ and\ \citenamefont
  {Uzan}}]{PhysRevLett.120.141101}%
  \BibitemOpen
  \bibfield  {author} {\bibinfo {author} {\bibfnamefont {J.}~\bibnamefont
  {Berg\'e}}, \bibinfo {author} {\bibfnamefont {P.}~\bibnamefont {Brax}},
  \bibinfo {author} {\bibfnamefont {G.}~\bibnamefont {M\'etris}}, \bibinfo
  {author} {\bibfnamefont {M.}~\bibnamefont {Pernot-Borr\`as}}, \bibinfo
  {author} {\bibfnamefont {P.}~\bibnamefont {Touboul}}, \ and\ \bibinfo
  {author} {\bibfnamefont {J.-P.}\ \bibnamefont {Uzan}},\ }\href {\doibase
  10.1103/PhysRevLett.120.141101} {\bibfield  {journal} {\bibinfo  {journal}
  {Phys. Rev. Lett.}\ }\textbf {\bibinfo {volume} {120}},\ \bibinfo {pages}
  {141101} (\bibinfo {year} {2018})}\BibitemShut {NoStop}%
\bibitem [{\citenamefont {Svelto}(2010)}]{Svelto2011PrinciplesOL}%
  \BibitemOpen
  \bibfield  {author} {\bibinfo {author} {\bibfnamefont {O.}~\bibnamefont
  {Svelto}},\ }\enquote {\bibinfo {title} {Properties of laser beams},}\ in\
  \href {\doibase 10.1007/978-1-4419-1302-9} {\emph {\bibinfo {booktitle}
  {Principles of Lasers}}}\ (\bibinfo  {publisher} {Springer New York, NY},\
  \bibinfo {year} {2010})\ p.\ \bibinfo {pages} {153}\BibitemShut {NoStop}%
\bibitem [{\citenamefont {Kogelnik}\ and\ \citenamefont
  {Li}(1966)}]{Kogelnik:66}%
  \BibitemOpen
  \bibfield  {author} {\bibinfo {author} {\bibfnamefont {H.}~\bibnamefont
  {Kogelnik}}\ and\ \bibinfo {author} {\bibfnamefont {T.}~\bibnamefont {Li}},\
  }\href {\doibase 10.1364/AO.5.001550} {\bibfield  {journal} {\bibinfo
  {journal} {Appl. Opt.}\ }\textbf {\bibinfo {volume} {5}},\ \bibinfo {pages}
  {1550} (\bibinfo {year} {1966})}\BibitemShut {NoStop}%
\end{thebibliography}%

\section*{APPENDIX: LIGHT FIELD CALCULATION AND MEASUREMENT NOISE OPTIMIZATION}\label{light}

\textit{Optical Calculation.} The light emitted from the incident fiber is assumed to be Gaussian, taking  the light propagation direction as the z-axis, the incident Gaussian light  intensity distribution at waist can be written as \cite{Svelto2011PrinciplesOL}:
 $$I_1 (r)=I_0 \mathrm{exp}\left(-\frac{2r^2}{\omega_{01}^2}\right)$$

And the waist radius of incident Gaussian beam is $\omega_{01}$, which satisfies relation:
$$\omega_{01}=\sqrt{\frac{a_0^2 \lambda^2}{\lambda^2+\pi^2 a_0^2 tan^2 \alpha}}$$
where $a_0$ is the radius of fiber core, and sin$\alpha$ = N.A, N.A. is the numerical aperture of the fiber. In there $a_0=5\mu$m and N.A.=0.13 for single-mode fiber. The incident optical power is:
$$P_{\mathrm{in}}=\int_0^{\infty} I_1 (r) 2 \pi rdr=\frac{\pi}{2} \omega_{01}^2 I_0$$

The response of the light to the micro-sphere is calculated using the standard optical ABCD ray matrix \cite{Kogelnik:66}. Under the par-axial approximation, the transmission matrix $\mathbf{T}$ is:
$$\mathbf{T}=\begin{pmatrix}
	A & B \\ C & D 
\end{pmatrix}$$

which has the equation:
\begin{equation*}
\begin{pmatrix}
	r_f \\ \theta_f
\end{pmatrix}
=
\mathbf{T}
\begin{pmatrix}
	r_i \\ \theta_i
\end{pmatrix}
\end{equation*}

In calculating the transmission matrix $\mathbf{T}$, we neglected the reflection of light at the interface and the absorption in the micro-sphere. Here A, B, C, D are
$$A=\frac{2}{n}-1,B=\frac{2R}{n},C=\frac{1-n}{n}\frac{2}{n},D=\frac{2}{n}-1,\beta_0=\frac{\lambda}{\pi \omega_{01}^2}$$
with the parameters $\lambda=1550 nm$, n=1.45, the we get the $d_2$ and $\omega_{02}$ satisfy
$$d_2=\frac{AC/\beta_0^2+ACd_1^2+ADd_1+BCd_1+BD}{C^2 /\beta_0^2+C^2 d_1^2+2CDd_1+D^2}$$
$$\omega_{02}=\omega_{01}\sqrt{(A+Cd_2 )^2+\beta_0^2(Ad_1+B+Cd_1 d_2+Dd_2 )^2}$$
$d_2$ and $\omega_{02}$ are functions of $d_1$, choose a suitable $d_1$ so that $\omega_{02}\approx a_0$. 
The coupling efficiency $\Gamma$, of the laser beam and the single-mode optical fiber can be written as:
$$\Gamma=\Gamma_0 \mathrm{exp}\bigg(-\Gamma_0\cdot \frac{x_{\mathrm{fib}}^2}{2} (\frac{1}{\omega_{02}^2} +\frac{1}{a_0^2})\bigg),
\Gamma_0=\frac{4\omega_{02}^2a_0^2}{(\omega_{02}^2+a_0^2 )^2}$$

%$$\Gamma_0=\frac{4\omega_{02}^2a_0^2}{(\omega_{02}^2+a_0^2 )^2} $$

\begin{figure}
	\centering
	\includegraphics[width=0.95\columnwidth]{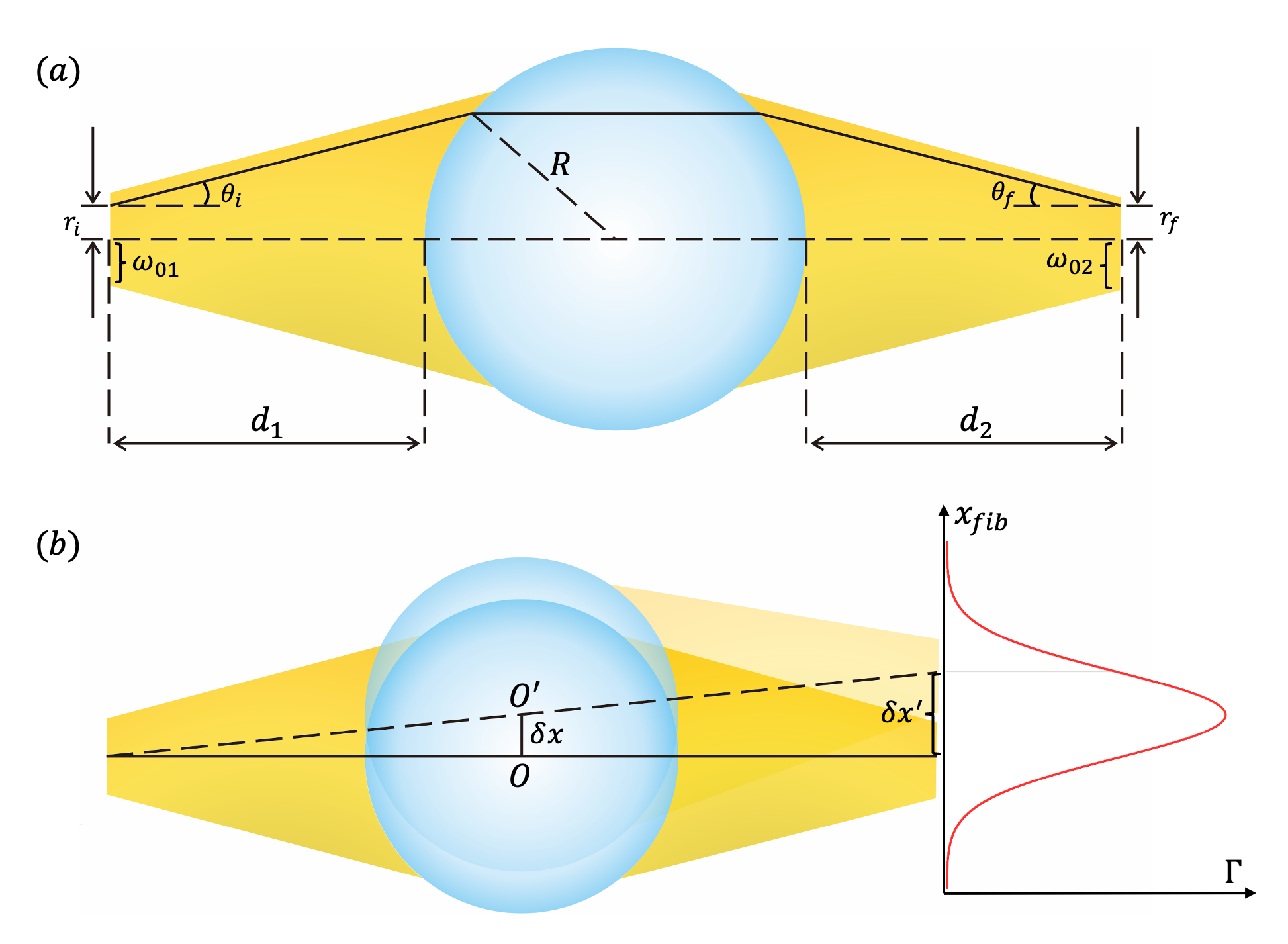}
	\caption{(a) Optical ray of the laser. $\theta_i$ ($\theta_f$) and $r_i$ 	($r_f$) are used to characterize the optical ray coming from incident 	fiber and reaching the detection fiber, $d_1$ ( $d_2$) are the distance 	between the incident fiber (detection fiber) and the optical axis, $R$ is 	the radius of the micro-sphere. (b) Dependence of the light field 	distribution with the microsphere position. The position of the image on 	the incident fiber core $\delta x^{\prime}$ in x axis depends on the 
		position of the micro-sphere position $\delta x$. The transmission 	coefficient $\Gamma$ changes with $\delta x$.}\label{fig6}
\end{figure}

$x_{\mathrm{fib}}$ indicate the fiber shift from the $x$ direction, when $x_{\mathrm{fib}}=0$, $\Gamma=\Gamma_{max}=\Gamma_0$.
In the experiment, fix $x_{\mathrm{fib}}$  at the place where $\partial \Gamma/\partial x_{\mathrm{fib}}$  is the largest. As $x_{\mathrm{fib}}=2.51\mu m$ and $\Gamma(x_{\mathrm{fib}})=0.604$ in Fig.\ref{fig6}(b).

$\delta x$ is the displacement of the micro-sphere vertically to the optical axis (similar result for y direction), while $\delta x'$ is the projection on the incident fiber surface. Under par-axial approximation, $\delta x=\zeta \cdot \delta x^{\prime}$ for small displacement $\delta x$ of the micro-sphere, with the displacement magnification factor:
$$\zeta=\frac{d_1+d_2+2R}{d_1+R},
\varsigma=\frac{\partial \Gamma}{\partial x}=\frac{\partial \Gamma}{\partial x'} \cdot \frac{\partial x'}{\partial x}=\zeta \cdot \frac{\partial \Gamma}{\partial x'}$$

\textit{Measurement Noise.} The relationship between the average power $P$ and the photon number $N$ is:
$$N_{\mathrm{in}}=\frac{P_{\mathrm{in}}T_{\mathrm{mea}}}{\hbar \omega_{\mathrm{op}}},
N_{\mathrm{dec}}=\frac{P_{\mathrm{dec}}T_{\mathrm{mea}}}{\hbar \omega_{\mathrm{op}}}$$
where $\omega_{\mathrm{op}}$ is the light frequency. The photons satisfy the Poisson distribution and the corresponding photon number fluctuation is $\delta N_{\mathrm{in}}=\sqrt{N_{\mathrm{in}}}$ and $\delta N_{\mathrm{dec}}=\sqrt{N_{\mathrm{dec}}}$. Such fluctuation brings a imprecise detection noise of displacement $\delta x_{\mathrm{imp}}$:
\begin{equation*}
	\begin{aligned}
\delta x_{\mathrm{imp}}&=\frac{\partial x}{\partial \Gamma}\sqrt{
	\left(\frac{\partial \Gamma}{\partial N_{\mathrm{in}}}\delta N_{\mathrm{in}} \right)^2+
	\left(\frac{\partial \Gamma}{\partial N_{\mathrm{dec}}}\delta N_{\mathrm{dec}} \right)^2} \\
&=\frac{1}{\varsigma} \sqrt{\frac{\Gamma+\Gamma^2}{N_{\mathrm{in}}}}
\end{aligned}
\end{equation*}

Thus the power density of displacement noise is:
$$S_{\mathrm{xx}}^{\mathrm{imp}}=\frac{1}{\varsigma^2}\frac{(\Gamma+\Gamma^2)\hbar \omega_{\mathrm{op}}}{P_{\mathrm{in}}}$$

On the other hand, the photon passes through the micro-sphere which changes the direction and therefore generated a back-action force $\delta f_{\mathrm{ba}}$ with the strength also proportional to the fluctuation of the incident photon $\delta N_{\mathrm{in}}$. The back-action force $\delta f_{\mathrm{ba}}$ can be written as:
$$\delta f_{\mathrm{ba}}=\sqrt{N_{\mathrm{in}}}\hbar \Delta k /T_{\mathrm{mea}}$$
where $\Delta k$ is the change of the wave vector.

Here we suppose that the direction of light wave vector is along the direction of the Gaussian light wavefront, and the probability of photon appearing is proportional to the intensity of Gaussian light. $\Delta k$ is the average change of light wave vector pass through the micro-sphere. It is calculated by $\sqrt{(\Delta k_{\mathrm{in}})^2+(\Delta k_{\mathrm{out}})^2}$, where $\Delta k_{\mathrm{in}}$ is the average light wave vector go to the micro-sphere, $\Delta k_{\mathrm{out}}$ is the average light wave vector go out of the micro-sphere. We obtain
\begin{equation*}
	\begin{aligned}
(\Delta k)^2=&k^2 \beta   \\
=&k^2 \int_0^{\infty} \frac{k^2 r^3}{k^2 r^2 +((1-\frac{z_r^2}{z_l^2})\frac{kR^2}{2\rho(z_l)}+\frac{z_r}{z}-k\rho(z_l))^2}  \cdot \\ &\frac{1}{\omega_1^2(z_l)} \mathrm{exp}\left(-\frac{2r^2}{\omega_1^2(z_l)}\right)dr
\end{aligned}
\end{equation*}
where $k=\omega_{\mathrm{op}}/c$, $z_l=d_1+R-\sqrt{R^2-r^2}$, $\omega_1(z_l)=\omega_{01}\sqrt{1+(z_l /z_r)^2}$, $z_r=2 \pi \omega_{01}^2 / \lambda$ and $\rho(z_l)=z_r (z_l/z_r +z_r/z_l)$.

The power density of back-action noise is thus: 
$$S_{\mathrm{ff}}^{\mathrm{ba}}=\frac{P_{\mathrm{in}} \hbar \omega_{\mathrm{op}} \beta}{c^2}$$
and the product of imprecision noise and back-action noise is:
$$S_{\mathrm{xx}}^{\mathrm{imp}}\cdot S_{\mathrm{ff}}^{\mathrm{ba}}=\frac{1}{\varsigma^2}  (\Gamma+\Gamma^2 ) 
(\omega_{\mathrm{op}} /c)^2 \beta^2 \hbar^2$$
The quantum efficiency of the measurement is defined as:
$$\eta=\frac{\varsigma}{4(\Gamma+\Gamma^2)\beta k^2}$$
where $\eta$ = 1 corresponding standard quantum limit (SQL). The total measurement noise is
$$S_{\mathrm{aa}}^{\mathrm{mea}} (\omega)=\frac{S_{\mathrm{xx}}^{\mathrm{imp}}}{|\chi_{\mathrm{\scriptscriptstyle m}}(\omega,\omega_0)|^2} 
+\frac{S_{\mathrm{ff}}^{\mathrm{ba}}}{m^2}$$

$S_{\mathrm{aa}}^{\mathrm{mea}}$ is minimized by tuning the incident laser power $P_{\mathrm{in}}$ under the product constraint of the imprecision noise and backaction noise. The optimized power is:
$$P_{\mathrm{opt}} (\omega,\omega_0 )=\sqrt{\frac{\Gamma+\Gamma^2}{\beta}}  
\frac{m c}{\varsigma|\chi_{\mathrm{\scriptscriptstyle m}}(\omega,\omega_0)|}$$
with the minimised total acceleration measurement noise as:
$$S_{\mathrm{aa,min}}^{\mathrm{mea}}=\frac{2\hbar \omega_{\mathrm{op}} }{m\varsigma c |\chi_{\mathrm{\scriptscriptstyle m}}(\omega,\omega_0)|}  \sqrt{\beta(\Gamma+\Gamma^2 )} $$

And in order to simplify the experiment process, we choose $P_{\mathrm{in}} =P_{\mathrm{opt}} (\omega_0,\omega_0)$, with the optimized  acceleration measurement noise at this time:
$$S_{\mathrm{aa,opt}}^{\mathrm{mea}}=\frac{\hbar \omega_{\mathrm{op}} 
	\sqrt{\beta(\Gamma+\Gamma^2 )}}{m\varsigma c \gamma \omega_0} \cdot 
\left(\frac{1}{	|\chi_{\mathrm{\scriptscriptstyle m}}(\omega,\omega_0)|^2}+\gamma^2 \omega_0^2 \right)$$

\end{document}